\documentclass[amssymb,aps]{revtex4-2}
\usepackage{graphicx}
\usepackage{enumerate}
\usepackage{amssymb}
\usepackage{amsmath}
\usepackage{amsfonts}
\usepackage{color}
\usepackage{mathrsfs}

\begin{document}
\title{Quantum optomechanics of lossy bodies: general approach and structured squeezed vacuum effects}

\author{A. Ciattoni$^1$}
\email{alessandro.ciattoni@spin.cnr.it}
\affiliation{$^1$CNR-SPIN, c/o Dip.to di Scienze Fisiche e Chimiche, Via Vetoio, 67100 Coppito (L'Aquila), Italy}
\date{\today}

\begin{abstract}
We investigate the overall optomechanical force experienced by a macroscopic lossy object in free space under external quantum illumination. To this end, utilizing the Modified Langevin Noise Formalism (MLNF), we derive the time-averaged expectation value of the Maxwell stress tensor for a non-equilibrium scenario in which the incoming scattering field is prepared in an arbitrary mixed quantum state, while the medium-assisted field is maintained in local thermal equilibrium. In the limit of full radiation-matter thermal equilibrium, our expression exactly recovers the well-known fluctuation-dissipation relation governing the Casimir effect, and, under coherent illumination, it yields the standard classical radiation pressure. We demonstrate that by driving the scattering field with an anisotropic, multimode squeezed vacuum state, the spatial profile of the electromagnetic quantum fluctuations can be engineered to exhibit broken rotational symmetry, thereby inducing a purely quantum mechanical force acting on the object. Such mechanical interaction is generated in the strict absence of a mean field, $\langle\hat{\mathbf{E}}\rangle=0$, and its non-classical nature is evidenced by its reliance on second-order field correlations $\langle\hat{\mathbf{E}}^2\rangle$, unlike classical optical radiation pressure governed by the squared mean field $\langle\hat{\mathbf{E}}\rangle^2$. Applying this exact formulation to a homogeneous lossy sphere, we demonstrate the experimental feasibility of the effect using realistic material parameters and optical estimations. Ultimately, we establish a general formalism for macroscopic quantum optomechanics that operates beyond the constraints of thermal equilibrium, enabling the prediction of regimes where the purely quantum force circumvents classical mean fields and shot noise while preserving the object's macroscopic quantum coherence.
\end{abstract}

\maketitle

\section{Introduction}
The macroscopic manifestations of quantum vacuum fluctuations represent a fundamental intersection between quantum field theory and macroscopic physics. A primary example is the Casimir force, an attractive mechanical interaction between neutral, polarizable bodies arising from the confinement of the electromagnetic zero-point energy \cite{Casim1948}. The generalization of this effect to arbitrary macroscopic lossy media was first accomplished by Lifshitz \cite{Lifsh1956}. Relying on fluctuational electrodynamics, Lifshitz's original derivation required the evaluation of Maxwell stress tensors driven by fluctuating sources obeying the Fluctuation-Dissipation Theorem (FDT) \cite{Rytov1989}. Later, the development of the Langevin noise formalism (LNF), also referred to as Macroscopic Quantum Electrodynamics (MQED), provided a systematic framework \cite{Gruna1996, Grunb1996, Schee1998, Dungg1998, Schee2008} to derive such fluctuational forces. By quantizing the electromagnetic field in the presence of dispersive and absorbing media via the introduction of continuous bosonic heat baths, LNF allows the Casimir-Lifshitz force to be derived directly from the zero-temperature ground state or the thermal equilibrium state of the coupled light-matter system \cite{Philb2011,Buhma2012}. However, standard LNF formulations typically focus on the medium-assisted field generated by quantized dipolar sources distributed throughout the volume of the absorbing medium. While adequate for systems in global thermal equilibrium, where the incoming external radiation is balanced with the medium emission, this approach often lacks the explicit inclusion of an independent scattering sector, and is inherently forced to model the incident field via limiting procedures of damped waves with vanishing damping \cite{Grunc1996, Knoll1999}, a procedure that becomes analytically demanding for non-planar geometries. An exact treatment of the scattering modes becomes necessary when moving beyond global thermal equilibrium. In recent years, theoretical and experimental studies have investigated non-equilibrium Casimir forces driven by temperature gradients, such as bodies maintained at different temperatures or objects immersed in an environment with a mismatched thermal bath \cite{Antez2005, Kruge2011, Messi2011, Antez2008, Bimon2009}. In these thermal non-equilibrium scenarios, the mechanical force is modified by the directional flow of thermal photons. Because these studies rely on thermal statistical mixtures (Bose-Einstein distributions) to populate the scattering modes, purely quantum non-equilibrium states remain largely unexplored.

Concurrently, the field of quantum optomechanics \cite{Aspel2014, Kippe2008, Meyst2013} has explored the quantum limits of mechanical systems interacting with tailored electromagnetic fields \cite{Genes2008, Marqu2007}. However, the macroscopic mechanical drive in these setups is conventionally achieved via a strong classical coherent field, generating a radiation pressure proportional to the squared mean field $\langle \hat{\mathbf{E}} \rangle^2$, often within the restricted geometry of 1D optical cavities \cite{Lawww1995, Fabre1994, Manci1994}. Significantly, the presence of such an intense mean field ($\langle \hat{\mathbf{E}} \rangle \neq 0$) inherently limits the quantum regime: it acts as a linear amplifier for vacuum fluctuations, amplifying radiation pressure shot noise \cite{Clerk2010, Aspel2014}, and its scattering subjects the mechanical oscillator to rapid spatial decoherence, effectively degrading macroscopic quantum superpositions \cite{Jooss1985, Schlo2007, Romer2011}. To mitigate this amplified shot noise, squeezed light is routinely injected into these setups; however, it almost invariably acts as a mere quantum perturbation atop the macroscopic, classical carrier field (e.g., a high-power laser) \cite{Caves1981, Clark2017, Tseee2019}. Consequently, engineering electromagnetic quantum fluctuations to produce a macroscopic, directional force in 3D free space—driven solely by time-averaged second-order field correlations $\langle \hat{\mathbf{E}}^2 \rangle$ in the strict absence of a classical mean field ($\langle \hat{\mathbf{E}} \rangle = 0$)—remains a fundamental open challenge. Crucially, operating in this purely quantum regime would inherently circumvent radiation pressure shot noise and macroscopic spatial decoherence. Describing this scenario from first principles requires handling arbitrary quantum illumination interacting with realistic, finite-size macroscopic lossy objects, a requirement precisely fulfilled by the modified Langevin noise formalism (MLNF) \cite{DiSte2001,Dreze2017,Dorie2019,Naaaa2023}, which specifically enables an explicit partitioning of the global Fock space into two orthogonal sectors: the scattering sector, spanned by the scattering-polariton operators, and the medium-assisted sector, generated by the electric and magnetic medium-polariton operators. As explicitly demonstrated in Refs.\cite{Ciatt2024, Ciatt2026}, the MLNF possesses a solid and canonical theoretical foundation, being the exact second-quantized version of the macroscopic electromagnetism quantum theory introduced in Ref.\cite{Philb2010}. Furthermore, owing to its capability to handle the scattering modes, the MLNF has recently been exploited to model the interaction of quantum emitters with dispersive objects \cite{Miana2025, Mianb2025, Miano2026} and to develop a general approach to quantum optical scattering by finite-size lossy objects in vacuum \cite{Ciata2025, Ciatb2025}.

In this work, we present a comprehensive theoretical description of the optomechanical force exerted on a macroscopic lossy object in 3D free space under arbitrary external quantum illumination. Leveraging the MLNF, we derive the time-averaged Maxwell stress tensor for the general non-equilibrium dynamics wherein the scattering polaritons are described by an arbitrary and time-evolving density operator, while the medium electric and magnetic polaritons are maintained in local thermodynamic equilibrium. As a necessary consistency check, we consider the limit of global radiation-matter thermal equilibrium, showing that the combined contributions of the two field sectors naturally recover the fluctuation-dissipation relation and the standard Lifshitz theory of Casimir forces. We then specialize our general formalism to the scenario in which the incident field is prepared in an anisotropic, multimode squeezed vacuum state, thereby modifying the spatial distribution of the electromagnetic quantum fluctuations. Such tailoring of the quantum field breaks the rotational symmetry of the vacuum’s momentum flux, generating a net mechanical force on the object. A distinguishing feature of this interaction is that it occurs in the strict absence of a mean field ($\langle \hat{\mathbf{E}} \rangle = 0$). While traditional optomechanics relies on the radiation pressure of coherent fields—typically governed by the squared mean field $\langle \hat{\mathbf{E}} \rangle^2$—the force investigated here originates solely from the structured second-order field correlations $\langle \hat{\mathbf{E}}^2 \rangle$. By analyzing the spatial distribution of these correlations, we demonstrate how quantum fluctuations can be harnessed to exert directional pressure on a macroscopic body without the assistance of any classical carrier. To extract analytical insights and assess experimental feasibility, we apply the theory to a homogeneous lossy sphere. This exact analysis highlights the interplay between the object's classical radiation pressure cross-section and the spatial distribution of the squeezing parameters, demonstrating the directional nature of the force and providing realistic magnitude estimations for the purely quantum mechanical manipulation of macroscopic objects. Our results establish a general framework for macroscopic quantum optomechanics capable of describing purely quantum regimes where the mechanical force arises solely from the spatial control of structured quantum fluctuations, independently of classical drives and thermal gradients.

The paper is organized as follows. In Sec.II, we introduce the theoretical model based on the MLNF, defining the exact dyadic field operators and the fundamental integral relations. In Sec.III, we define the general non-equilibrium scenario and derive the time-averaged Maxwell stress tensor along with the exact decomposition of the spectral correlation dyadic. In Sec.IV, we evaluate the spectral correlation dyadic for specific external field preparations, analyzing the cases of thermal radiation, coherent illumination, and multimode squeezed vacuum. In Sec.V, we derive the exact analytical expression for the optomechanical force under squeezed illumination, highlighting the competition between the active quantum drive and the passive thermal recoil. In Sec.VI, we apply the macroscopic formalism to a homogeneous lossy sphere to extract analytical insights via exact Mie optical cross-sections and provide realistic magnitude estimations to assess the experimental feasibility of the effect. Finally, in Sec.VII, we outline our conclusions.

\section{Modified Langevin noise formalism}
In this section, we briefly recall the main features of the Modified Langevin Noise Formalism (MLNF) \cite{Ciatt2024, Ciatt2026}, whose foundational details are outlined in Appendix A. The framework provides the description of the quantum electromagnetic field in the presence of an arbitrary finite-size, inhomogeneous, dispersive, and lossy magnetodielectric object, whose optical response is encoded in the complex electric permittivity $\varepsilon_\omega(\mathbf{r})$ and magnetic permeability $\mu_\omega(\mathbf{r})$ (both equal to $1$ in the surrounding vacuum). Within this framework, if the Schr\"odinger picture is chosen, the Hamiltonian operator $\hat{H}$ and the electric field operator ${\mathbf{\hat E}}$ are given by
\begin{eqnarray} \label{HamEle}
\hat H &=& \int {d\omega } \;\hbar \omega \left[ {\int {do_{\mathbf{n}} } {\mathbf{\hat g}}_{\omega s}^\dag  \left( {\mathbf{n}} \right) \cdot {\mathbf{\hat g}}_{\omega s} \left( {\mathbf{n}} \right) + \sum\limits_{\nu} {\int {d^3 {\mathbf{r}}\,} {\mathbf{\hat f}}_{\omega \nu }^\dag  \left( {\mathbf{r}} \right) \cdot {\mathbf{\hat f}}_{\omega \nu } \left( {\mathbf{r}} \right)} } \right]  , \nonumber \\ 
  {\mathbf{\hat E}}\left( {\mathbf{r}} \right) &=& \int{d\omega } \left[ {\int {do_{\mathbf{n}} } {\mathcal F}_{\omega s} \left( {\left. {\mathbf{r}} \right|{\mathbf{n}}} \right) \cdot {\mathbf{\hat g}}_{\omega s} \left( {\mathbf{n}} \right) + \sum\limits_{\nu} {\int {d^3 {\mathbf{r}}'\,} {\mathcal G}_{\omega \nu } \left( {\left. {\mathbf{r}} \right|{\mathbf{r}}'} \right) \cdot {\mathbf{\hat f}}_{\omega \nu } \left( {{\mathbf{r}}'} \right)} } \right] + {\rm H.c.}.
\end{eqnarray}
These expressions are formulated in terms of the scattering polariton operators ${\mathbf{\hat g}}_{\omega s} \left( {\mathbf{n}} \right)$ and the medium-assisted polariton operators ${\mathbf{\hat f}}_{\omega \nu } \left( {\mathbf{r}} \right)$ ($\nu \in \{e, m\}$), which satisfy the bosonic commutation relations
\begin{align} \label{MLNF_Commutators_Main}
\left[ {{\mathbf{\hat g}}_{\omega s} \left( {\mathbf{n}} \right),{\mathbf{\hat g}}_{\omega 's}^\dag  \left( {{\mathbf{n}}'} \right)} \right] &= \delta \left( {\omega  - \omega '} \right)\delta \left( {o_{\mathbf{n}}  - o_{{\mathbf{n}}'} } \right){\mathcal I}_{\mathbf{n}}, &
\left[ {{\mathbf{\hat f}}_{\omega \nu } \left( {\mathbf{r}} \right),{\mathbf{\hat f}}_{\omega '\nu '}^\dag  \left( {{\mathbf{r}}'} \right)} \right] &= \delta \left( {\omega  - \omega '} \right)\delta _{\nu \nu '} \delta \left( {{\mathbf{r}} - {\mathbf{r}}'} \right){\mathcal I},
\end{align}
where all other commutators vanish. Here, ${\mathcal I}$ is the identity dyadic, ${\mathcal I}_{\mathbf{n}}$ is the dyadic projector onto the plane transverse to the direction ${\mathbf{n}}$, and $\delta (o_{\mathbf{n}} - o_{{\mathbf{n}}'})$ is the angular delta function. Furthermore, the dyadic kernel ${\mathcal F}_{\omega s} ({\mathbf{r}}|{\mathbf{n}})$ appearing in the electric field operator is related to the two scattering modes produced by plane waves incoming from the direction ${\mathbf{n}}$. Conversely, the kernel ${\mathcal G}_{\omega \nu } ({\mathbf{r}}|{\mathbf{r}}')$ is related to the dyadic Green's function of classical electrodynamics, but it identically vanishes for source points ${\mathbf{r}}'$ located outside the volume of the lossy medium. The association of these specific dyadic kernels with their respective polaritonic operators directly justifies the nomenclature adopted for the latter. Crucially, these dyadic kernels satisfy the fundamental integral relation
\begin{equation} \label{Integral_Relation}
\int {do_{\mathbf{n}} } {\mathcal F}_{\omega s} \left( {\left. {\mathbf{r}} \right|{\mathbf{n}}} \right) \cdot {\mathcal F}_{\omega s}^{T*} \left( {\left. {{\mathbf{r}}'} \right|{\mathbf{n}}} \right) + \int {d^3 {\mathbf{s}}} \sum\limits_\nu  {{\mathcal G}_{\omega \nu } \left( {\left. {\mathbf{r}} \right|{\mathbf{s}}} \right) \cdot {\mathcal G}_{\omega \nu }^{T*} \left( {\left. {{\mathbf{r}}'} \right|{\mathbf{s}}} \right)}  = \frac{{\hbar k_\omega ^2 }}{{\pi \varepsilon _0 }}{\mathop{\rm Im}\nolimits} \left[ {{\mathcal G}_\omega  \left( {\left. {\mathbf{r}} \right|{\mathbf{r}}'} \right)} \right].
\end{equation}
It should be noted that in the standard LNF, the imaginary part of the dyadic Green's function ${{\mathcal G}_\omega  \left( {\left. {\mathbf{r}} \right|{\mathbf{r}}'} \right)}$ is reconstructed solely through the medium-assisted contribution (second term in the LHS). While omitting the scattering contribution (first term on the LHS) is physically consistent for spatially unbounded media (the specific regime for which the standard LNF is formulated), Eq.\eqref{Integral_Relation} explicitly demonstrates that for a bounded lossy object, the inclusion of these scattering modes is strictly required. This additional term guarantees the completeness of the quantization scheme and ensures the rigorous fulfillment of the fluctuation-dissipation theorem. 

The diagonal form of the Hamiltonian in Eq.\eqref{HamEle} demonstrates that the quanta associated with the polaritonic operators (the scattering and medium-assisted polaritons) are indeed the true elementary excitations of the system, effectively mapping the complex interaction between the radiation and the lossy body onto a set of independent bosonic modes. Consequently, the global Fock space of the electromagnetic field is the direct sum of three mutually orthogonal sectors: the scattering ($s$) sector, representing incoming radiation excitations, and the electric ($e$) and magnetic ($m$) medium-assisted sectors, associated with the quantized dipolar sources within the object's volume $V$ arising from its dissipative nature.

It is worth emphasizing that the MLNF rigorously incorporates absorption and dispersion without ad-hoc approximations. Although the Kramers-Kronig relations dictate that the absolute vacuum is the only strictly transparent medium, it is physically instructive to artificially consider the transparent limit (${\mathop{\rm Im}\nolimits} [\varepsilon_\omega(\mathbf{r})] \to 0$, ${\mathop{\rm Im}\nolimits} [\mu_\omega(\mathbf{r})] \to 0$) in which the electric and magnetic dyadic kernels ${\mathcal G}_{\omega e}$ and ${\mathcal G}_{\omega m}$ identically vanish. This leads to the complete collapse of the medium-assisted sector, effectively decoupling the medium polaritons ${\mathbf{\hat f}}_{\omega \nu }$ from the field dynamics. Consequently, Eq.\eqref{Integral_Relation} reduces to the completeness relation for the modes associated with the scattering polaritons; these remain the sole elementary excitations of the system and correspond exactly to the standard photons of the macroscopic quantization theory of Glauber and Lewenstein \cite{Glaub1991}.

\section{Momentum flow around a thermalized macroscopic object under arbitrary quantum illumination}
We consider a general non-equilibrium scenario where the incident radiation field is specified by an arbitrary density operator $\hat{\rho}_s(t)$ acting on the $s$-sector and evolving according to the Liouville-von Neumann equation $i\hbar d\hat{\rho}_s / dt = [\hat{H}_s, \hat{\rho}_s(t)]$, where $\hat{H}_s$ is the scattering part of the Hamiltonian in Eqs.\eqref{HamEle}. Conversely, the object is maintained in local thermodynamic equilibrium at temperature $T_{em}$ and is described by the density operator $\hat{\rho}_{em} = e^{-\beta_{em} \hat{H}_{em}} / {\rm Tr}(e^{-\beta_{em} \hat{H}_{em}})$ acting on the medium-assisted ($e$ and $m$) sectors, where $\beta_{em} = (k_B T_{em})^{-1}$. Assuming statistical independence between the $s$ and $em$ sectors, the total density operator is $\hat{\rho}_s(t) \hat{\rho}_{em}$, which enables the calculation of the quantum-statistical expectation value of an arbitrary operator $\hat{O}$ as 
\begin{equation}
\langle \hat{O} \rangle (t) = \text{Tr} \left[ \hat{\rho}_s(t) \hat{\rho}_{em} \hat{O} \right],
\end{equation}
where $\text{Tr}$ denotes the operator trace performed over the full Fock space. The time-averaged expectation value of the operator is then defined accordingly as
\begin{equation}
\langle\langle \hat{O} \rangle\rangle = \lim_{T \to +\infty} \frac{1}{T} \int_{-T/2}^{T/2} dt \, \langle \hat{O} \rangle (t).
\end{equation}
Owing to the factorization of the total density operator, the bosonic nature of the medium polariton operators ${{\mathbf{\hat f}}_{\omega \nu } \left( {\mathbf{r}} \right)}$ allows one to rigorously prove that
\begin{align} \label{thermal_averages}
\langle {\mathbf{\hat f}}_{\omega \nu} ({\mathbf{r}}) \rangle &= 0, & \langle {\mathbf{\hat f}}_{\omega \nu}^\dag ({\mathbf{r}}) {\mathbf{\hat f}}_{\omega' \nu'} ({\mathbf{r}}') \rangle &= n_\omega (\beta_{em}) \delta(\omega - \omega') \delta_{\nu \nu'} \delta({\mathbf{r}} - {\mathbf{r}}') \mathcal{I},
\end{align}
where $n_\omega (\beta) = (e^{\beta \hbar \omega} - 1)^{-1}$ is the Bose-Einstein distribution, as physically expected since the medium assisted field (i.e., the body) is maintained in local thermal equilibrium. Note that, since $\hat{\rho}_{em}$ is strictly stationary, these quantum-statistical averages inherently coincide with their time-averaged counterparts.

The net momentum flow at a vacuum point outside the object is quantified by the time-averaged expectation value $\langle\langle {\hat {\mathcal T}} \rangle\rangle$ of the Maxwell stress dyadic operator
\begin{equation} \label{Maxwell_Tensor}
\hat{\mathcal{T}} = \varepsilon_0 \left[ \hat{\mathbf{E}} \hat{\mathbf{E}} - \frac{1}{2} (\hat{\mathbf{E}} \cdot \hat{\mathbf{E}}) \mathcal{I} \right] + \frac{1}{\mu_0} \left[ \hat{\mathbf{B}} \hat{\mathbf{B}} - \frac{1}{2} (\hat{\mathbf{B}} \cdot \hat{\mathbf{B}}) \mathcal{I} \right],
\end{equation}
whose use, in addition to established electrodynamic principles, is here rigorously justified by the momentum conservation analysis of Ref.\cite{Philb2011}. Indeed, such an analysis is rooted in the canonical quantization of macroscopic electromagnetism \cite{Philb2010}, of which the MLNF employed in this work constitutes the exact second-quantized version \cite{Ciatt2024, Ciatt2026}. As shown in Appendix B, the time-averaged expectation value of the Maxwell stress dyadic operator is
\begin{equation} \label{AvT}
\langle\langle {\hat {\mathcal T}} ({\mathbf r})\rangle\rangle  = \varepsilon _0 \int {d\omega } \left\{ {\left[ {{\mathcal C}_\omega  \left( {{\mathbf{r}}\left| {{\mathbf{r}}'} \right.} \right) - \frac{1}{{k_\omega ^2 }}\nabla _{\mathbf{r}}  \times {\mathcal C}_\omega  \left( {{\mathbf{r}}\left| {{\mathbf{r}}'} \right.} \right) \times \mathord{\buildrel{\lower3pt\hbox{$\scriptscriptstyle\leftarrow$}} 
\over \nabla } _{{\mathbf{r}}'} } \right] - \frac{1}{2} {\rm tr} \left[ {{\mathcal C}_\omega  \left( {{\mathbf{r}}\left| {{\mathbf{r}}'} \right.} \right) - \frac{1}{{k_\omega ^2 }}\nabla _{\mathbf{r}}  \times {\mathcal C}_\omega  \left( {{\mathbf{r}}\left| {{\mathbf{r}}'} \right.} \right) \times \mathord{\buildrel{\lower3pt\hbox{$\scriptscriptstyle\leftarrow$}} 
\over \nabla } _{{\mathbf{r}}'} } \right]{\mathcal I}} \right\}_{{\mathbf{r}}' \to {\mathbf{r}}} 
\end{equation}
where $\text{tr}$ denotes the dyadic trace, whereas 
\begin{eqnarray} \label{Com}
 {\mathcal C}_\omega  \left( {{\mathbf{r}}\left| {{\mathbf{r}}'} \right.} \right) &=& \frac{{\hbar k_\omega ^2 }}{{\pi \varepsilon _0 }}{\mathop{\rm Im}\nolimits} \left[ {{\mathcal G}_\omega  \left( {{\mathbf{r}}\left| {{\mathbf{r}}'} \right.} \right)} \right] + 2{\mathop{\rm Re}\nolimits} \int {do_{\mathbf{n}} } \int {do_{{\mathbf{n}}'} } {\mathcal F}_{\omega s} \left( {\left. {\mathbf{r}} \right|{\mathbf{n}}} \right) \cdot \left[ {\int {d\omega '} \left\langle\left\langle {{\mathbf{\hat g}}_{\omega s}^\dag  \left( {\mathbf{n}} \right){\mathbf{\hat g}}_{\omega 's} \left( {{\mathbf{n}}'} \right)} \right\rangle\right\rangle } \right]^* \cdot {\mathcal F}_{\omega s}^{T*} \left( {\left. {{\mathbf{r}}'} \right|{\mathbf{n}}'} \right) \nonumber \\ 
  &+& 2n_\omega  \left( {\beta _{em} } \right){\mathop{\rm Re}\nolimits} \int {d^3 {\mathbf{s}}\,} \sum\limits_\nu  {{\mathcal G}_{\omega \nu } \left( {\left. {\mathbf{r}} \right|{\mathbf{s}}} \right) \cdot {\mathcal G}_{\omega \nu }^{T*} \left( {\left. {{\mathbf{r}}'} \right|{\mathbf{s}}} \right)}  
\end{eqnarray}
is the spectral correlation dyadic. Equation (\ref{Com}) provides a rigorous decomposition of ${\mathcal C}_\omega ({\mathbf{r}}|{\mathbf{r}}')$ into its fundamental physical contributions. Because the adopted formalism retains all terms and introduces no \emph{a priori} approximations, the equation reflects the exact partition of the field fluctuations into three fundamental physical contributions. The first term, proportional to ${\mathop{\rm Im}\nolimits} [{\mathcal G}_\omega ({\mathbf{r}}|{\mathbf{r}}')]$, represents the intrinsic electromagnetic zero-point fluctuations. Consistent with the zero-temperature fluctuation-dissipation theorem, the local density of electromagnetic states is entirely encoded in the imaginary part of the system's dyadic Green's function. Dynamically, this term accounts for the Casimir-Lifshitz forces in the absolute vacuum, reproducing standard literature expressions for the Casimir force between two objects at zero temperature. The second term describes the contribution of external electromagnetic fields prepared in arbitrary quantum states. The time-averaged expectation value $\langle\langle {\mathbf{\hat g}}_{\omega s}^\dag ({\mathbf{n}}){\mathbf{\hat g}}_{\omega 's} ({\mathbf{n}}') \rangle\rangle$ dictates how the statistical and coherence properties of the incident field (e.g., classical coherent states or non-classical squeezed light) spatially propagate through the dyadic kernel ${\mathcal F}_{\omega s}$. This addend quantifies externally induced optical forces, including gradient forces and radiation pressure, while accounting for the quantum nature of the source. The third term, proportional to the Bose-Einstein distribution $n_\omega (\beta _{em})$, isolates the thermal contribution from the bath of medium-polaritons characterizing the object's thermodynamic equilibrium. Vanishing at absolute zero, this term quantifies internal thermal fluctuations and constitutes the exact thermal contribution of the medium-assisted field (medium-polaritons) to the Casimir-Lifshitz force.

To further analyze the electromagnetic fluctuations, the zero-point contribution can be expanded. Applying the fundamental integral relation of Eq.(\ref{Integral_Relation}) to Eq.(\ref{Com}) yields
\begin{eqnarray} \label{ComExpanded}
 {\mathcal C}_\omega  \left( {{\mathbf{r}}\left| {{\mathbf{r}}'} \right.} \right) &=& {\mathop{\rm Re}\nolimits} \int {do_{\mathbf{n}} } \int {do_{{\mathbf{n}}'} } {\mathcal F}_{\omega s} \left( {\left. {\mathbf{r}} \right|{\mathbf{n}}} \right) \cdot \left[ {{\mathcal I}_{\mathbf{n}} \delta \left( {o_{\mathbf{n}}  - o_{{\mathbf{n}}'} } \right) + 2\int {d\omega '} \left\langle\left\langle {{\mathbf{\hat g}}_{\omega s}^\dag  \left( {\mathbf{n}} \right){\mathbf{\hat g}}_{\omega 's} \left( {{\mathbf{n}}'} \right)} \right\rangle\right\rangle } \right]^* \cdot {\mathcal F}_{\omega s}^{T*} \left( {\left. {{\mathbf{r}}'} \right|{\mathbf{n}}'} \right) \nonumber \\ 
  &+& \coth \left( {\frac{{\beta _{em} \hbar \omega }}{2}} \right){\mathop{\rm Re}\nolimits} \int {d^3 {\mathbf{s}}\,} \sum\limits_\nu  {{\mathcal G}_{\omega \nu } \left( {\left. {\mathbf{r}} \right|{\mathbf{s}}} \right) \cdot {\mathcal G}_{\omega \nu }^{T*} \left( {\left. {{\mathbf{r}}'} \right|{\mathbf{s}}} \right)}.
\end{eqnarray}
This representation clarifies the spatial and statistical decoupling of the correlator into two independent fluctuation channels, a feature that naturally facilitates the rigorous analytical and numerical evaluation of the ensuing quantum interactions. The first line groups the external radiative contributions. It shows that the bare radiative vacuum, represented by ${\mathcal I}_{\mathbf{n}} \delta (o_{\mathbf{n}} - o_{{\mathbf{n}}'})$, and the excitations of the incident quantum field, encoded in $\langle\langle {\mathbf{\hat g}}_{\omega s}^\dag ({\mathbf{n}}){\mathbf{\hat g}}_{\omega 's} ({\mathbf{n}}') \rangle\rangle$, are structurally equivalent. Both components propagate and diffract around the object through the identical spatial dyadic kernel ${\mathcal F}_{\omega s}$, allowing the correlation properties of the total external field to be tailored, at least in principle, by macroscopic quantum states. Simultaneously, the second line combines the internal material vacuum, arising from local dissipation, with the thermal medium-polariton bath. The hyperbolic cotangent factor, mathematically equivalent to $1 + 2n_\omega(\beta_{em})$, confirms that the internal degrees of freedom strictly obey the local fluctuation-dissipation theorem at thermal equilibrium, acting as distributed stochastic sources throughout the volume of the lossy object. This formulation ensures that any external manipulation of the incident field occurs without altering the thermodynamic consistency of the internal material fluctuations.

To further elucidate the exact decomposition in Eq.(\ref{ComExpanded}), it is instructive to revisit the transparent limit (${\mathop{\rm Im}\nolimits} [\varepsilon_\omega({\bf r})] \to 0$ and ${\mathop{\rm Im}\nolimits} [\mu_\omega ({\bf r})] \to 0$). Since the medium-assisted dyadic kernels ${\mathcal G}_{\omega \nu}$ identically vanish (see Sec. II A), the second line of Eq.(\ref{ComExpanded}) collapses to zero. This reduction yields two main physical implications. First, it dictates a strict thermodynamic decoupling: the spectral correlation dyadic becomes entirely independent of the internal temperature parameter $\beta_{em}$. Consistent with Kirchhoff's law, a non-absorbing object cannot emit thermal photons, making it thermally decoupled from the surrounding vacuum regardless of its internal temperature $T_{em}$. Second, since the real parts of the material response remain finite (${\mathop{\rm Re}\nolimits} [\varepsilon_\omega] \neq 1$), the modal dyadic ${\mathcal F}_{\omega s}$ still accounts for refraction and diffraction. Consequently, any optomechanical interaction in this regime arises exclusively from the elastic scattering of the bare zero-point fluctuations and the incoming quantum excitations (encoded in the first line of Eq.(\ref{ComExpanded})). The object experiences a purely conservative momentum transfer, entirely devoid of the thermal recoil that would otherwise stem from internal material dissipation.

\section{Spectral correlation dyadic for thermal, coherent, and squeezed illumination}
Having established the general expression for the spectral correlation dyadic in Eq.(\ref{ComExpanded}), we now evaluate it for three distinct physical preparations of the external scattering field. Specifically, we analyze the cases of thermal radiation, coherent illumination, and a multimode squeezed vacuum. By systematically specifying the scattering density operator $\hat{\rho}_s(t)$ for each scenario, we demonstrate how the statistical and quantum coherence properties of the incident radiation uniquely shape the spatial distribution of the electromagnetic fluctuations. This analysis provides the necessary foundation for understanding how different external field states drive the optomechanical interaction, transitioning from macroscopic thermal gradients and deterministic classical fields to purely quantum, fluctuation-induced forces.

\subsection{Thermal radiation} 
As a first application of the general formalism, we consider the scenario where the incident scattering polaritons are in thermal equilibrium at a temperature $T_s$, which is generally different from the internal temperature $T_{em}$ of the lossy object. In this regime, the scattering density operator is time independent and it is given by $\hat{\rho}_s = e^{-\beta_s \hat{H}_s} / \text{Tr}(e^{-\beta_s \hat{H}_s})$, where $\hat{H}_s$ is the scattering part of the Hamiltonian. This implies that the normally ordered quantum-statistical correlator of the scattering polariton operators evaluates to $\langle \hat{\mathbf{g}}_{\omega s}^\dagger(\mathbf{n}) \hat{\mathbf{g}}_{\omega' s}(\mathbf{n}') \rangle = n_\omega(\beta_s) \delta(\omega - \omega') \delta(o_{\mathbf{n}} - o_{\mathbf{n}'}) \mathcal{I}_{\mathbf{n}}$, where $n_\omega(\beta_s)$ is the Bose-Einstein distribution at temperature $T_s$. Since this state is strictly stationary, its time-averaged counterpart trivially coincides with it, i.e., $\langle\langle \hat{\mathbf{g}}_{\omega s}^\dagger(\mathbf{n}) \hat{\mathbf{g}}_{\omega' s}(\mathbf{n}') \rangle\rangle = \langle \hat{\mathbf{g}}_{\omega s}^\dagger(\mathbf{n}) \hat{\mathbf{g}}_{\omega' s}(\mathbf{n}') \rangle$. Substituting this expression into Eq.(\ref{ComExpanded}), the dyadic correlator takes the form
\begin{equation} \label{Com_Therm_rad}
\mathcal{C}_\omega (\mathbf{r}|\mathbf{r}') = \text{Re} \left[ \coth \left( \frac{\beta_s \hbar \omega}{2} \right) \int do_{\mathbf{n}} \mathcal{F}_{\omega s} (\mathbf{r}|\mathbf{n}) \cdot \mathcal{F}_{\omega s}^{T*} (\mathbf{r}'|\mathbf{n}) + \coth \left( \frac{\beta_{em} \hbar \omega}{2} \right) \int d^3\mathbf{s} \sum_\nu \mathcal{G}_{\omega \nu} (\mathbf{r}|\mathbf{s}) \cdot \mathcal{G}_{\omega \nu}^{T*} (\mathbf{r}'|\mathbf{s}) \right].
\end{equation}
In this non-equilibrium thermal regime, the physical origin of the ensuing optomechanical interactions resides entirely in the electromagnetic fluctuations, since the macroscopic mean field strictly vanishes,
\begin{equation}
\langle \hat{\mathbf{E}}(\mathbf{r}) \rangle = \text{Tr} \left[ \hat{\rho}_s \hat{\rho}_{em} \hat{\mathbf{E}}(\mathbf{r}) \right] = 0.
\end{equation}
Equation (\ref{Com_Therm_rad}) illustrates that while the formal symmetry between the s-sector and medium-assisted sector contributions is completely preserved, the numerical difference in their temperatures ($T_s \neq T_{em}$) produces a non-trivial spatial structure in the correlation dyadic. Accordingly, the ensuing optomechanical force can be dynamically tuned via the macroscopic temperature difference $\Delta T = T_s - T_{em}$. This result directly reproduces the non-equilibrium Casimir forces previously investigated via macroscopic fluctuational electrodynamics and scattering-matrix approaches \cite{Antez2005, Antez2008, Bimon2009, Kruge2011, Messi2011}. While these established treatments typically require phenomenologically partitioning the coupled radiation-matter system to separately evaluate the classical stochastic currents within the medium and the background radiation from infinity, the MLNF circumvents this procedural complexity. Rather, the MLNF natively encodes these non-equilibrium dynamics through the distinct thermal populations of its orthogonal polaritonic sectors, yielding a physically transparent framework that entirely bypasses the need for supplementary boundary-matching or scattering-matrix techniques.

In the limit of global thermal equilibrium, where the external radiation and the macroscopic body share the same temperature, $T = T_s = T_{em}$ (and thus $\beta = \beta_s = \beta_{em}$), the hyperbolic cotangent factors in Eq.(\ref{Com_Therm_rad}) become identical and can be factored out. By virtue of the fundamental integral relation in Eq.(\ref{Integral_Relation}), the angular scattering integral and the spatial medium-assisted integral exactly recombine to reconstruct the imaginary part of the dyadic Green's function, yielding
\begin{equation}
\mathcal{C}_\omega (\mathbf{r}|\mathbf{r}') = \coth \left( \frac{\beta \hbar \omega}{2} \right) \frac{\hbar k_\omega^2}{\pi \varepsilon_0} \text{Im} [\mathcal{G}_\omega (\mathbf{r}|\mathbf{r}')].
\end{equation}
This result perfectly coincides with the predictions of the fluctuation-dissipation theorem at finite temperature. It demonstrates that the MLNF recovers the standard Lifshitz theory \cite{Lifsh1956}, offering a direct and physically transparent derivation of the Casimir-Lifshitz force that avoids the complexities of macroscopic fluctuational electrodynamics. Furthermore, while the standard LNF \cite{Philb2011, Buhma2012} also yields this equilibrium expression by formally treating the surrounding vacuum as an unbounded material medium with vanishingly small absorption, the MLNF reconstructs it without requiring such asymptotic limiting procedures, natively recombining the physically distinct contributions of the independent external scattering modes and the medium-assisted contribution. A strictly analogous condition emerges in the quantum electrodynamics of two-level emitters interacting with macroscopic dispersive bodies. As recently demonstrated \cite{Miana2025, Mianb2025, Miano2026}, the MLNF description of the atomic spontaneous emission and the corresponding field correlations rigorously reduces to the standard LNF results solely when the scattering and the medium-assisted sectors are in full thermal equilibrium. This further highlights the fundamental necessity of an independent scattering sector to properly capture any quantum optical phenomenon operating out of global thermal equilibrium.

\subsection{Coherent illumination}
To evaluate the optomechanical response under a classical driving field, such as a continuous-wave laser illumination, we consider the scenario where the external scattering polaritons are prepared in a multimode coherent state. For this pure state, the density operator is $\hat{\rho}_s(t) = | \{ e^{-i\omega t} \boldsymbol{\alpha}_\omega(\mathbf{n}) \} \rangle \langle \{ e^{-i\omega t} \boldsymbol{\alpha}_\omega(\mathbf{n}) \} |$, where the state is formally defined by the displacement operator acting on the vacuum $|0\rangle$
\begin{equation}
\left| \left\{ e^{-i\omega t} \boldsymbol{\alpha}_\omega(\mathbf{n}) \right\} \right\rangle = \exp \left\{ \int d\omega \int do_{\mathbf{n}} \left[ e^{-i\omega t} \boldsymbol{\alpha}_\omega(\mathbf{n}) \cdot \hat{\mathbf{g}}_{\omega s}^\dagger(\mathbf{n}) - e^{i\omega t} \boldsymbol{\alpha}_\omega^*(\mathbf{n}) \cdot \hat{\mathbf{g}}_{\omega s}(\mathbf{n}) \right] \right\} |0\rangle,
\end{equation}
where the complex amplitude $\boldsymbol{\alpha}_\omega(\mathbf{n})$ is a vector field defined on the unit sphere, constrained by the transversality condition $\mathbf{n} \cdot \boldsymbol{\alpha}_\omega(\mathbf{n}) = 0$. The normally ordered expectation value of the scattering operators is evaluated via a long-time average to extract the stationary spectral contribution
\begin{equation}
\langle\langle \hat{\mathbf{g}}_{\omega s}^\dagger(\mathbf{n}) \hat{\mathbf{g}}_{\omega' s}(\mathbf{n}') \rangle\rangle = \lim_{T \to +\infty} \frac{1}{T} \int_{-T/2}^{T/2} dt \, \langle \hat{\mathbf{g}}_{\omega s}^\dagger(\mathbf{n}) \hat{\mathbf{g}}_{\omega' s}(\mathbf{n}') \rangle = \lim_{T \to +\infty} \frac{1}{T} \int_{-T/2}^{T/2} dt \, e^{i(\omega - \omega')t} \boldsymbol{\alpha}_\omega^*(\mathbf{n}) \boldsymbol{\alpha}_{\omega'}(\mathbf{n}').
\end{equation}
Assuming the coherent illumination is strictly stationary, its spectral amplitude is sharply peaked at a driving frequency $\omega_0$, taking the form $\boldsymbol{\alpha}_\omega(\mathbf{n}) = \delta(\omega - \omega_0)\mathbf{A}(\mathbf{n})$, where $\mathbf{A}(\mathbf{n})$ specifies the angular distribution of the incident field. Under this condition, the correlator evaluates to $\langle\langle \hat{\mathbf{g}}_{\omega s}^\dagger(\mathbf{n}) \hat{\mathbf{g}}_{\omega' s}(\mathbf{n}') \rangle\rangle = \delta(\omega - \omega_0)\delta(\omega' - \omega_0) \mathbf{A}^*(\mathbf{n})\mathbf{A}(\mathbf{n}')$. Substituting this expectation value into the general expression of Eq.(\ref{ComExpanded}), the spectral correlation dyadic becomes
\begin{eqnarray} \label{Com_Cohe}
 \mathcal{C}_\omega (\mathbf{r}|\mathbf{r}') &=& \text{Re} \left[ \int do_{\mathbf{n}} \mathcal{F}_{\omega s} (\mathbf{r}|\mathbf{n}) \cdot \mathcal{F}_{\omega s}^{T*} (\mathbf{r}'|\mathbf{n}) + \coth \left( \frac{\beta_{em} \hbar \omega}{2} \right) \int d^3\mathbf{s} \sum_\nu \mathcal{G}_{\omega \nu} (\mathbf{r}|\mathbf{s}) \cdot \mathcal{G}_{\omega \nu}^{T*} (\mathbf{r}'|\mathbf{s}) \right] \nonumber \\ 
  &+& 2\delta (\omega - \omega_0) \text{Re} \left[ \mathbf{E}_{\omega_0}^{(cl)}(\mathbf{r}) \mathbf{E}_{\omega_0}^{(cl)*} (\mathbf{r}') \right], 
\end{eqnarray}
where we have defined the macroscopic coherent field profile $\mathbf{E}_{\omega_0}^{(cl)}(\mathbf{r}) = \int do_{\mathbf{n}} \mathcal{F}_{\omega_0 s}(\mathbf{r}|\mathbf{n}) \cdot \mathbf{A}(\mathbf{n})$. This profile represents the complex amplitude of the monochromatic field $\mathbf{E}^{(cl)}(\mathbf{r},t) = 2\text{Re} [ e^{-i\omega_0 t} \mathbf{E}_{\omega_0}^{(cl)}(\mathbf{r}) ]$ describing the classical scattering produced by the object under an incident radiation specified, in the far past, by the angular amplitudes $\mathbf{A}(\mathbf{n})$. As expected for a coherent state, this classical field exactly coincides with the instantaneous quantum expectation value of the electric field operator, i.e.,
\begin{equation} \label{Coe-MeanField}
\langle {\bf{\hat E}}\left( {\bf{r}} \right) \rangle = {\rm Tr} \left[ {\hat \rho _s \left( t \right)\hat \rho _{em} {\bf{\hat E}}\left( {\bf{r}} \right)} \right] = {\bf{E}}^{\left( {cl} \right)} \left( {{\bf{r}},t} \right).
\end{equation}
When inserted into the Maxwell stress dyadic of Eq.(\ref{AvT}), the bilinear term $\mathbf{E}_{\omega_0}^{(cl)}(\mathbf{r})\mathbf{E}_{\omega_0}^{(cl)*}(\mathbf{r}')$ of Eq.(\ref{Com_Cohe}) generates a macroscopic momentum flow governed by the local classical field intensity $|\mathbf{E}_{\omega_0}^{(cl)}(\mathbf{r})|^2$, which exactly coincides with the standard classical radiation pressure exerted by the incident field. Thus, Eq.(\ref{Com_Cohe}) reveals that optomechanical actions in this regime comprise two fundamentally different contributions. The first is a stochastic, fluctuation-driven component arising from the scattering-sector vacuum and the medium-assisted thermal equilibrium, states where the mean electric field strictly vanishes. The second is a deterministic radiation pressure governed by the time average of the squared classical field. It is precisely this reliance on a non-vanishing mean field that allows the second deterministic contribution to inherently survive in the classical limit.

\subsection{Multimode squeezed vacuum illumination}
To evaluate the optomechanical response under a non-classical driving field, we consider the scenario where the external scattering polaritons are prepared in a multimode squeezed vacuum state. For this pure state, the density operator is $\hat{\rho}_s(t) = | \{ \zeta_\omega(\mathbf{n})e^{-i2\omega t} \} \rangle \langle \{ \zeta_\omega(\mathbf{n})e^{-i2\omega t} \} |$, where the state vector is generated by the application of the multimode squeeze operator onto the vacuum $|0\rangle$
\begin{equation}
\left| \left\{ \zeta_\omega(\mathbf{n})e^{-i2\omega t} \right\} \right\rangle = \exp \left\{ \frac{1}{2} \int d\omega \int do_{\mathbf{n}} \left[ \zeta_\omega^*(\mathbf{n})e^{i2\omega t} \hat{\mathbf{g}}_{\omega s}(\mathbf{n}) \cdot \hat{\mathbf{g}}_{\omega s}(\mathbf{n}) - \zeta_\omega(\mathbf{n})e^{-i2\omega t} \hat{\mathbf{g}}_{\omega s}^\dagger(\mathbf{n}) \cdot \hat{\mathbf{g}}_{\omega s}^\dagger(\mathbf{n}) \right] \right\} |0\rangle,
\end{equation}
with the complex squeezing parameter defined as $\zeta_\omega(\mathbf{n}) = r_\omega(\mathbf{n})e^{i\phi_\omega(\mathbf{n})}$. The quantum-statistical expectation value of the normally ordered scattering operators evaluates to the transverse identity dyadic $\mathcal{I}_{\mathbf{n}}$ scaled by the characteristic stationary photon number distribution of the squeezed vacuum
\begin{equation}
\langle \hat{\mathbf{g}}_{\omega s}^\dagger(\mathbf{n}) \hat{\mathbf{g}}_{\omega' s}(\mathbf{n}') \rangle = \left\langle \left\{ \zeta_\omega(\mathbf{n})e^{-i2\omega t} \right\} \right| \hat{\mathbf{g}}_{\omega s}^\dagger(\mathbf{n}) \hat{\mathbf{g}}_{\omega' s}(\mathbf{n}') \left| \left\{ \zeta_\omega(\mathbf{n})e^{-i2\omega t} \right\} \right\rangle = \sinh^2 [r_\omega(\mathbf{n})] \delta(\omega - \omega') \delta(o_{\mathbf{n}} - o_{\mathbf{n}'}) \mathcal{I}_{\mathbf{n}}
\end{equation}
so that the long-time average trivially preserves the corresponding spectral contribution yielding $\langle\langle \hat{\mathbf{g}}_{\omega s}^\dagger(\mathbf{n}) \hat{\mathbf{g}}_{\omega' s}(\mathbf{n}') \rangle\rangle = \sinh^2 [r_\omega(\mathbf{n})] \delta(\omega - \omega') \delta(o_{\mathbf{n}} - o_{\mathbf{n}'}) \mathcal{I}_{\mathbf{n}}$. Substituting this expectation value into Eq.(\ref{ComExpanded}) yields the explicit form of the spectral correlation dyadic for the squeezed illumination
\begin{eqnarray} \label{Com_Squeezed}
\mathcal{C}_\omega (\mathbf{r}|\mathbf{r}') &=& \text{Re} \int do_{\mathbf{n}} \left\{ 1 + 2\sinh^2 [r_\omega(\mathbf{n})] \right\} \mathcal{F}_{\omega s} (\mathbf{r}|\mathbf{n}) \cdot \mathcal{F}_{\omega s}^{T*} (\mathbf{r}'|\mathbf{n}) \nonumber \\ 
&+& \text{Re} \int d^3\mathbf{s} \sum_\nu \left\{ 1 + 2n_\omega(\beta_{em}) \right\} \mathcal{G}_{\omega \nu} (\mathbf{r}|\mathbf{s}) \cdot \mathcal{G}_{\omega \nu}^{T*} (\mathbf{r}'|\mathbf{s}).
\end{eqnarray}
This formulation reveals that, under non-classical squeezed illumination, a fundamental physical symmetry emerges between the scattering and medium-assisted sectors. The squeezed state effectively acts as an anisotropic thermal environment whose effective temperature and angular distribution are fully tunable via the squeezing parameter $r_\omega(\mathbf{n})$. Its spectral contribution pairs a unit term for zero-point fluctuations with the stationary correlated photon population $\sinh^2 [r_\omega(\mathbf{n})]$. This bipartite structure perfectly mirrors the medium-assisted sector, which identically combines the vacuum unit term with the stationary Bose-Einstein population $n_\omega(\beta_{em})$ of the medium polaritons. This physical symmetry reflects a rigorous dynamical equivalence, since both contributions are fundamentally driven by pure field fluctuations. Indeed, under squeezed vacuum illumination, the macroscopic mean field identically vanishes at all times,
\begin{equation} \label{squ_mean_fie}
\langle \hat{\mathbf{E}}(\mathbf{r}) \rangle = \text{Tr} \left[ \hat{\rho}_s(t) \hat{\rho}_{em} \hat{\mathbf{E}}(\mathbf{r}) \right] = 0.
\end{equation} 
To compare this regime with the previously discussed scenarios, we note that while the optomechanical interactions in the two-temperature thermal case are similarly driven entirely by fluctuations, their tunability is strictly limited to the macroscopic temperature difference. The squeezed vacuum, instead, provides a substantially richer set of degrees of freedom, allowing the optomechanical response to be actively tailored via the full spectral and angular profile of the squeezing parameter $r_\omega(\mathbf{n})$. Furthermore, in sharp contrast to coherent illumination, which is dominated by the deterministic radiation pressure of a non-vanishing mean field, the structured squeezed vacuum generates a purely non-classical mechanical force entirely devoid of any classical carrier.

\section{Quantum force on a lossy body under squeezed illumination}
In this section, we evaluate the net optomechanical force exerted on a single macroscopic object immersed in an external multimode squeezed vacuum. While squeezed light is conventionally employed in optomechanics as a noise-reduction technique alongside a dominant classical drive, here the macroscopic mechanical interaction originates solely from the spatial structuring of pure vacuum fluctuations, entirely devoid of any non-vanishing classical mean field. This purely quantum force inherently avoids the radiation pressure shot noise and spatial decoherence associated with intense coherent fields, delineating a distinct physical mechanism that remains unaddressed in the current literature. Furthermore, as discussed in Sec.IIIC, this non-classical illumination scheme affords extensive spatial and spectral control over the optomechanical response. It provides a significantly richer degree of tunability than the macroscopic gradients of a two-temperature thermal environment, while strictly preserving the zero-mean-field condition that coherent illumination inherently violates.

The net optomechanical force $\mathbf{F}$ experienced by the object is obtained by integrating the time-averaged expectation value of the Maxwell stress dyadic, given by Eq.(\ref{AvT}), over any closed surface enclosing the body. Since the object is isolated in vacuum, it is analytically convenient to evaluate this surface integral over a sphere of infinite radius, yielding
\begin{equation} \label{Force}
\mathbf{F} = \lim_{r \to \infty} r^2 \int do_{\mathbf{r}} \, \mathbf{u}_r \cdot \langle\langle \hat{\mathcal{T}}(\mathbf{r}) \rangle\rangle,
\end{equation}
where $\mathbf{u}_r = \mathbf{r}/r$ is the radial unit vector and $do_{\mathbf{r}}$ is the solid angle element. In this expression, the momentum flux is rigorously evaluated by substituting the spectral correlation dyadic $\mathcal{C}_\omega (\mathbf{r}|\mathbf{r}')$ specifically derived for the squeezed vacuum illumination in Eq.(\ref{Com_Squeezed}). As extensively detailed in Appendix C, carrying out this surface integration requires separating the total momentum flux into its scattering and medium-assisted contributions. By evaluating the far-field asymptotic behavior of the modal dyadics and exploiting the fundamental integral relations of the MLNF to exactly handle the medium-assisted fields, the integral can be solved analytically. This rigorous derivation ultimately yields the exact expression for the total optomechanical force:
\begin{equation} \label{Eq_Force_Final}
{\bf{F}} = {\bf{F}}_{sca} \left[ {\sinh ^2 \left[ {r_\omega  \left( {\bf{n}} \right)} \right]} \right] - {\bf{F}}_{sca} \left[ {n_\omega  \left( {\beta _{em} } \right)} \right]
\end{equation}
where
\begin{equation} \label{F_sca}
{\bf{F}}_{sca} \left[ {w_\omega  \left( {\bf{n}} \right)} \right] = \int {d\omega } \frac{{\hbar k_\omega ^3 }}{{8\pi ^3 }}\int {do_{\bf{n}} w_\omega  \left( {\bf{n}} \right)} \left\{{\bf{n}} {\frac{{4\pi }}{{k_\omega  }}{\mathop{\rm Im}\nolimits} \left[ {{\rm tr}{\cal S}_\omega  \left( {{\bf{n}}\left| {\bf{n}} \right.} \right)} \right] - \int {do_{\bf{m}} } {\bf{m}}\;{\rm tr}\left[ {{\cal S}_\omega  \left( {{\bf{m}}\left| {\bf{n}} \right.} \right) \cdot {\cal S}_\omega ^{T*} \left( {{\bf{m}}\left| {\bf{n}} \right.} \right)} \right]} \right\}.
\end{equation}
where ${\cal S}_\omega({\bf{m}}|{\bf{n}})$ is the scattering dyadic of classical electrodynamics, which describes the amplitude of the field scattered by the object into the outgoing direction ${\bf{m}}$ for an incident plane wave impinging from direction ${\bf{n}}$ (see Appendix A for a detailed discussion).

Equation (\ref{Eq_Force_Final}) constitutes a central result of this study. It expresses the exact macroscopic optomechanical force on a dissipative object as the strict competition between two physical agents: the active quantum drive and the passive medium-assisted thermodynamic reaction. Crucially, reflecting the fact that the squeezed vacuum and the thermal background share the same physical nature originating from field fluctuations, as established in Sec.III C, both the quantum squeezing parameter and the thermodynamic photon number enter the exact same universal radiation pressure functional ${\bf{F}}_{sca}[w_\omega({\bf{n}})]$. Consequently, the total force naturally emerges as a direct subtraction, unifying the external structured push and the internal thermal recoil into a single effective statistical weight $\Delta w_\omega({\bf{n}}) = \sinh^2[r_\omega({\bf{n}})] - n_\omega(\beta_{em})$. This exact formulation notably demonstrates the rigorous algebraic cancellation of the force exerted by the unperturbed zero-point vacuum, ensuring that the macroscopic dynamics are driven strictly by the active thermal and squeezed excitations, free from divergent background contributions.

The underlying mechanism of momentum transfer is entirely captured by the functional ${\bf{F}}_{sca}$, whose curly braces define the vectorial momentum transfer cross-section for a plane wave impinging from direction ${\bf{n}}$. The first term, proportional to ${\bf{n}}\,{\mathop{\rm Im}\nolimits}[{\rm tr}{\cal S}_\omega({\bf{n}}|{\bf{n}})]$, relies on the generalized optical theorem to quantify the total momentum flux extracted from the incident mode through both absorption and elastic scattering (extinction). The subtracted second term isolates the contribution of the elastically scattered light; within it, the trace evaluates the intensity summed over all polarization states along the outgoing direction ${\bf{m}}$, so that the integration over the solid angle $do_{\bf{m}}$ exactly yields the momentum re-radiated into the far-field. The exact difference between the total extracted flux and this re-radiated momentum yields the true radiation pressure vector, rigorously accounting for both direct momentum absorption and the recoil induced by the asymmetric redirection of the scattered light.

By evaluating this functional, the term ${\bf{F}}_{sca}[\sinh^2[r_\omega({\bf{n}})]]$ represents the active macroscopic push exerted by the externally structured squeezed vacuum, where $\sinh^2[r_\omega({\bf{n}})]$ is precisely the directional photon number density of the impinging quantum field. Conversely, the subtracted term ${\bf{F}}_{sca}[n_\omega(\beta_{em})]$ governs the thermal recoil induced by the object's own spontaneous emission. The explicit subtraction in Eq. (\ref{Eq_Force_Final}) carries a precise physical interpretation dictated by Kirchhoff's law of thermal radiation: the net momentum lost via anisotropic thermal emission is mathematically identical to the opposite of the radiation pressure the body would experience if illuminated by a perfectly isotropic thermal bath. This exact momentum balance provides, in principle, a versatile tuning mechanism: the net observable force can be actively controlled by simultaneously exploiting the engineering of the squeezed vacuum, such as its spectral properties and spatial directivity, and the thermodynamic temperature of the macroscopic body.

Finally, the structure of the functional ${\bf{F}}_{sca}$ reveals a fundamental geometric property concerning any arbitrary spectral weight $w_\omega$ that is isotropic, meaning it is independent of the incidence direction ${\bf{n}}$. In such cases, the generic functional ${\bf{F}}_{sca}[w_\omega]$ reduces to the unweighted solid-angle integration of the momentum transfer cross-section over all incident directions. If the macroscopic object possesses spatial inversion symmetry (e.g., a homogeneous sphere or a regular cylinder), this global angular integration identically vanishes. This mathematical property has two immediate physical consequences. First, if the externally applied squeezed vacuum were completely isotropic, it would exert exactly zero net force on a symmetric object. Second, by the very same principle, because the Bose-Einstein distribution $n_\omega(\beta_{em})$ is intrinsically isotropic, the thermal recoil functional ${\bf{F}}_{sca}[n_\omega]$ is rigorously zero for symmetric bodies: the spontaneous thermal emission is perfectly balanced in all opposite directions, yielding no net recoil regardless of the internal temperature. Therefore, the emergence of a net macroscopic force necessitates either an explicit breaking of spatial inversion symmetry in the material geometry, or a deliberately anisotropic quantum drive, where the engineered directivity of the squeezed vacuum ${\bf{F}}_{sca}[\sinh^2[r_\omega({\bf{n}})]]$ remains the exclusive active driver of the mechanical interaction.

To verify the physical consistency of Eq.(\ref{Eq_Force_Final}), we consider the macroscopic force in the transparent limit (${\mathop{\rm Im}\nolimits} [\varepsilon_\omega({\bf r})] \to 0$ and ${\mathop{\rm Im}\nolimits} [\mu_\omega ({\bf r})] \to 0$). For a non-absorbing object, the generalized optical theorem dictates that the extinction cross-section equals the elastic scattering cross-section, yielding $\frac{{4\pi }}{{k_\omega  }}{\mathop{\rm Im}\nolimits} \left[ {{\rm tr}{\cal S}_\omega  \left( {{\bf{n}}\left| {\bf{n}} \right.} \right)} \right] = \int {do_{\bf{m}} } {\rm tr}\left[ {{\cal S}_\omega  \left( {{\bf{m}}\left| {\bf{n}} \right.} \right) \cdot {\cal S}_\omega ^{T*} \left( {{\bf{m}}\left| {\bf{n}} \right.} \right)} \right]$. Substituting this identity into the radiation pressure functional we get
\begin{equation}
{\bf{F}}_s \left[ {w_\omega  \left( {\bf{n}} \right)} \right] = \int {d\omega } \frac{{\hbar k_\omega ^3 }}{{16\pi ^3 }}\int {do_{\bf{n}} } w_\omega  \left( {\bf{n}} \right)\int {do_{\bf{m}} } \left( {{\bf{n}} - {\bf{m}}} \right){\rm tr}\left[ {{\cal S}_\omega  \left( {{\bf{m}}\left| {\bf{n}} \right.} \right) \cdot {\cal S}_\omega ^{T*} \left( {{\bf{m}}\left| {\bf{n}} \right.} \right)} \right].
\end{equation}
In this expression, the vectorial factor $\left( {\bf{n}} - {\bf{m}} \right)$ explicitly quantifies the momentum transfer resulting from the elastic redirection of an incoming excitation from direction ${\bf{n}}$ into the outgoing direction ${\bf{m}}$. This firmly establishes that for a transparent medium, the optomechanical force arises exclusively from the geometrical scattering of the incident momentum, with no momentum being transferred to the internal degrees of freedom. Furthermore, by exploiting the reciprocity relation of the scattering dyadic given in Eq.\eqref{RecipSom}, it can be shown that the double angular integral $\int {do_{\bf{n}} } \int {do_{\bf{m}} } \left( {{\bf{n}} - {\bf{m}}} \right) {\rm tr} \left[ {{\cal S}_\omega  \left( {{\bf{m}}\left| {\bf{n}} \right.} \right) \cdot {\cal S}_\omega ^{T*} \left( {{\bf{m}}\left| {\bf{n}} \right.} \right)} \right] = 0 $ identically vanishes. Consequently, for any isotropic weight $w_\omega$, the pure-scattering functional disappears for an object of arbitrary shape, ${\bf{F}}_s \left[ {w_\omega  } \right] = 0$, effectively relaxing the requirement of spatial inversion symmetry. This geometric cancellation ensures that ${\mathbf{F}}_{sca}[n_\omega(\beta_{em})] = 0$ for asymmetric transparent bodies, consistent with the thermodynamic requirement that a non-absorbing medium cannot emit thermal radiation and thus experiences no thermal recoil. This perfectly agrees with the fact that the term ${\mathbf{F}}_{sca}[n_\omega(\beta_{em})]$ originates from the medium-assisted sector of the MLNF, which vanishes identically in the transparent limit (${\mathcal G}_{\omega \nu} \to 0$), as shown in Sec.IIA.

\section{Application to a homogeneous macroscopic sphere}
\subsection{Theoretical derivation for spherical targets}
To quantify the optomechanical interaction, we evaluate the exact force exerted on a nonmagnetic, homogeneous, isotropic lossy sphere of radius $a$. For a spherically symmetric object, the scattering properties are rotationally invariant, and the scattering dyadic ${\cal S}_\omega({\bf{m}}|{\bf{n}})$ is completely described by Mie theory \cite{Krist2016, Bornn2019, Bohre1983}. As detailed in Appendix D, this symmetry allows the tensorial trace properties of the scattering dyadic to be mapped directly onto the standard macroscopic scalar optical cross-sections. Specifically, due to rotational symmetry, the momentum transfer vector enclosed in the curly braces of the functional ${\bf{F}}_{sca}[w_\omega]$ defined in Eq.(\ref{F_sca}) must be strictly aligned with the incidence direction ${\bf{n}}$. By exploiting the integral trace relations for the extinction and asymmetry cross-sections summarized in Appendix D, this vectorial quantity exactly reduces to $2(\sigma_\omega^{ext} - \sigma_\omega^{asym}){\bf{n}} = 2\sigma_\omega^{pr}{\bf{n}}$, where $\sigma_\omega^{pr}$ is the classical radiation pressure cross-section. The general optomechanical functional for the homogeneous sphere therefore simplifies to
\begin{equation}
{\bf{F}}_{sca}[w_\omega({\bf{n}})] = \int d\omega \frac{\hbar k_\omega^3}{4\pi^3} \sigma_\omega^{pr} \int do_{\bf{n}} w_\omega({\bf{n}}) {\bf{n}},
\end{equation}
where the cross-section $\sigma_\omega^{pr}$ is explicitly evaluated via the Mie multipole expansion coefficients $a_{n\omega}$ and $b_{n\omega}$. Finally, the total macroscopic force on the sphere is obtained by substituting the quantum-statistical weights of the nonequilibrium fields into this simplified functional. Since the Bose-Einstein distribution $n_\omega(\beta_{em})$ characterizing the medium-assisted thermal bath is isotropic, its angular integration over the incidence direction vector ${\bf{n}}$ identically vanishes ($\int do_{\bf{n}} n_\omega(\beta_{em}) {\bf{n}} = 0$). This confirms that the sphere experiences zero thermal recoil, as required by its spatial inversion symmetry. The net quantum force is thus driven exclusively by the engineered anisotropy of the external squeezed vacuum, yielding
\begin{equation} \label{Force_Sphere_Mie}
{\bf{F}} = \int d\omega \frac{\hbar k_\omega^3}{4\pi^3} \sigma_\omega^{pr} \int do_{\bf{n}} \sinh^2 [r_\omega({\bf{n}})] {\bf{n}}.
\end{equation}
Equation (\ref{Force_Sphere_Mie}) demonstrates that the optomechanical force on a macroscopic sphere originates solely from the coupling between the material radiation pressure cross-section and the directional asymmetry of the quantum squeezing parameter $r_\omega({\bf{n}})$. Although $\sigma_\omega^{pr}$ is conventionally termed the classical radiation pressure cross-section, its appearance here does not imply that the force is driven by classical radiation pressure. With the mean field strictly vanishing (see Sec. IIIC), $\sigma_\omega^{pr}$ acts purely as a transfer function governing the momentum extracted from the second-order quantum fluctuations.

\begin{figure}
\centering
\includegraphics[width = 1\linewidth]{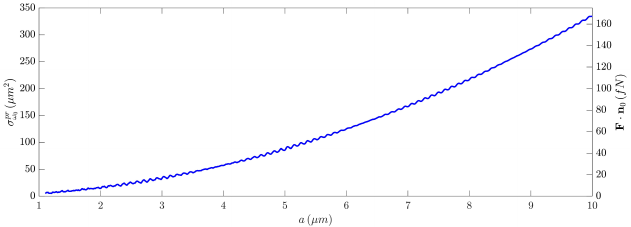}
\caption{Radiation pressure cross-section $\sigma_{\omega_0}^{pr}$ (left axis) and stationary nonclassical macroscopic force ${\bf F} \cdot {\bf n}_0$ (right axis) experienced by a homogeneous lossy sphere as a function of its radius $a$. The target material is modeled with a complex relative permittivity $\varepsilon = 12.11 + 0.1i$, representative of silicon microparticles, evaluated at the central vacuum wavelength $\lambda_0 = 1550 \text{ nm}$. The macroscopic force is computed assuming an external ultrabroadband squeezed vacuum illumination, yielding an equivalent nonclassical pressure $P_{sq} \simeq 0.5 \text{ fN}/\mu\text{m}^2$. The fine oscillatory structure superimposed on the purely geometric profile reflects the morphology-dependent multipolar Mie resonances characteristic of weakly dissipative high-index dielectrics.}
\label{Fig1}
\end{figure}

\subsection{Experimental feasibility and force magnitude estimation}

To quantify the achievable macroscopic force, we evaluate the integral in Eq.(\ref{Force_Sphere_Mie}) by noting that even for ultrabroadband squeezing in the telecom regime, the frequency spread $\Delta \omega$ remains a small fraction of the central carrier frequency $\omega_0$ (e.g., $\Delta \omega / \omega_0 \sim 10^{-2}$). Consequently, the complex material cross-section $\sigma_\omega^{pr}$ and the density-of-states factor $k_\omega^3$ vary slowly over the squeezed spectrum and can be robustly evaluated at $\omega_0$, thereby factoring out of the spectral integration. Furthermore, assuming the field is paraxially concentrated along the incidence axis ${\bf{n}}_0$, the force vector strictly aligns with ${\bf{n}}_0$. The exact functional evaluates to
\begin{equation} \label{Force_Estimation}
{\bf{F}} \simeq \sigma_{\omega_0}^{pr} \left[ \frac{\hbar k_{\omega_0}^3}{4\pi^3} \int d\omega \int do_{\bf{n}} \sinh^2 [r_\omega({\bf{n}})] \right] {\bf{n}}_0 = \sigma_{\omega_0}^{pr} P_{sq} {\bf{n}}_0,
\end{equation}
where we have introduced the equivalent quantum radiation pressure
\begin{equation} \label{P_sq_def}
P_{sq} = \frac{\hbar k_{\omega_0}^3}{4\pi^3} \sinh^2 (r_0) \Delta \omega_{eff} \Delta o_{eff},
\end{equation}
with $r_0$ being the peak squeezing parameter. Here, $\Delta \omega_{eff}$ and $\Delta o_{eff}$ are rigorously defined as the effective spectral bandwidth and effective solid angle of the stationary correlated photon distribution, respectively, absorbing any geometric factor arising from the specific nonlinear profile of $\sinh^2 [r_\omega({\bf{n}})]$. To provide a realistic estimation, we utilize state-of-the-art parameters for continuous-wave single-pass squeezing at a vacuum wavelength $\lambda_0 = 1550 \text{ nm}$. Recent advances in periodically poled lithium niobate (PPLN) waveguides allow for an effective ultrabroad bandwidth $\Delta \omega_{eff} \approx 2.5 \text{ THz}$ with a noise reduction of $6 \text{ dB}$ ($r_0 \simeq 0.69$) \cite{Kashi2020}. Assuming the correlated fluctuations are tightly focused by a high-numerical-aperture objective, yielding an effective solid angle $\Delta o_{eff} \approx 1 \text{ sr}$, the equivalent radiation pressure exerted exclusively by the quantum vacuum evaluates to $P_{sq} \simeq 0.5 \text{ fN}/\mu\text{m}^2$ ($5.0 \times 10^{-16} \text{ N}/\mu\text{m}^2$).

To maximize the optomechanical coupling, we consider a spherical target composed of a high-index lossy dielectric with a relative permittivity $\varepsilon = 12.11 + 0.1i$ evaluated at the telecommunication wavelength $\lambda_0 = 1550 \text{ nm}$. This specific parameter space is representative of silicon microparticles, which are widely employed in state-of-the-art Mie-resonant photonics to achieve strong optical confinement \cite{Fenol2019}. In Fig. 1, we report the exact calculation of the radiation pressure cross-section $\sigma_{\omega_0}^{pr}$ alongside the corresponding macroscopic stationary force ${\bf F} \cdot {\bf n}_0$ as a function of the sphere radius $a$. The strictly linear relationship between the two physical quantities is governed by the equivalent quantum radiation pressure $P_{sq} \simeq 0.5 \text{ fN}/\mu\text{m}^2$. Due to the moderate imaginary part of the silicon permittivity, the internal cavity resonances are not completely damped. This enables the manifestation of morphology-dependent multipolar Mie resonances (accurately resolved by retaining $200$ terms in the multipole series), which emerge as a fine oscillatory structure superimposed on the monotonically increasing quadratic profile that asymptotically approaches the geometric optics limit. For a macroscopic sphere with a radius $a = 10 \mu\text{m}$, the purely quantum radiation pressure yields a stationary nonclassical force exceeding $160 \text{ fN}$. Such a force magnitude is fundamentally independent of thermal background noise (as demonstrated in Sec.VA) and falls well within the high-fidelity detection capabilities of modern optomechanical force sensors, demonstrating that ultrabroadband squeezed vacuum fluctuations can induce a directly measurable macroscopic displacement.

\section{Conclusions}

In this work, we have established a comprehensive and first-principles framework for the quantum optomechanics of macroscopic dissipative bodies by employing the MLNF. This formalism provides an exact description of the electromagnetic momentum flow that overcomes the structural limitations inherent in existing theoretical approaches. Our derivation of the time-averaged Maxwell stress tensor rigorously unifies the contributions from the external scattering sector and the internal medium-assisted fluctuations, offering a universal platform to investigate mechanical forces under arbitrary quantum illumination and non-equilibrium thermal conditions. We have shown that in the limit of full radiation-matter thermal equilibrium, our approach recovers the standard Casimir-Lifshitz (and LNF) description, while for incident quantum coherent states, it successfully reproduces the classical radiation pressure results. This approach resolves the long-standing challenge of describing the momentum exchange between quantized fields and realistic, finite-size objects where both material loss and the independent scattering of incoming modes are physically indispensable. To illustrate the unique predictive power of this general formulation, we have investigated the notable illumination case of an anisotropic multimode squeezed vacuum. We have demonstrated that such non-classical illumination generates a macroscopic, directional mechanical force in the strict absence of a classical mean field ($\langle \hat{\mathbf{E}} \rangle = 0$). This result is particularly significant as it delineates a regime of purely fluctuational optomechanics that remains fundamentally inaccessible to traditional theoretical treatments; specifically, standard LNF typically lacks an independent scattering sector, while classical scattering theories are inherently incapable of accounting for the structured zero-point fluctuations within dissipative volumes. In this context, our approach predicts the active mechanical manipulation of macroscopic bodies through the spatial engineering of vacuum fluctuations, inherently circumventing the radiation pressure shot noise and the rapid spatial decoherence associated with intense coherent carriers. Finally, we have demonstrated the experimental feasibility of this fluctuation-driven momentum transfer by applying the framework to high-index silicon microspheres. The prediction of a stationary non-classical force exceeding $160$ fN for a target with a radius of $10 \mu\text{m}$ confirms that these effects are well within the high-fidelity detection capabilities of contemporary optomechanical sensors. By providing a rigorous foundation for macroscopic quantum optomechanics that operates independently of classical drives and thermal gradients, this work establishes a new paradigm for the control of macroscopic mechanical systems via structured quantum fluctuations.

\appendix

\section{Modified Langevin noise formalism}
We consider an arbitrary finite-size lossy object filling a spatial region $V$ in vacuum. In the frequency domain ($e^{-i \omega t}$), its inhomogeneous, isotropic magneto-dielectric response is described by the complex permittivity $\varepsilon _\omega ^{V} \left( {\mathbf{r}} \right)$ and permeability $\mu _\omega ^{V} \left( {\mathbf{r}} \right)$, which are holomorphic functions for ${\mathop{\rm Im}\nolimits} \left( \omega  \right) > 0$ to preserve causality. It is convenient to define the global piecewise responses ${\varepsilon _\omega  \left( {\mathbf{r}} \right)}$ and ${\mu _\omega  \left( {\mathbf{r}} \right)}$ as
\begin{align} \label{eps_mu}
\varepsilon _\omega  \left( {\mathbf{r}} \right) =& \left\{ {\begin{array}{*{20}l}
   {\varepsilon _\omega ^V \left( {\mathbf{r}} \right),} & {{\mathbf{r}} \in V}  \\
   {1,} & {{\mathbf{r}} \notin V}  \\
\end{array} } \right. &
\mu _\omega  \left( {\mathbf{r}} \right) =& \left\{ {\begin{array}{*{20}l}
   {\mu _\omega ^V \left( {\mathbf{r}} \right),} & {{\mathbf{r}} \in V}  \\
   {1,} & {{\mathbf{r}} \notin V}  \\
\end{array}} \right.
\end{align}
The essential classical electrodynamic tools for the MLNF are the modal dyadic ${\mathcal F}_\omega  \left( {\left. {\mathbf{r}} \right|{\mathbf{n}}} \right)$ and the dyadic Green's function ${{\mathcal G}_\omega  \left( {\left. {\mathbf{r}} \right|{\mathbf{r}}'} \right)}$, governed by the respective boundary-value problems
\begin{align} \label{GS}
\left[ {\left( {\nabla _{\mathbf{r}}  \times \frac{1}{{\mu _\omega  \left( {\mathbf{r}} \right)}}\nabla _{\mathbf{r}}  \times } \right) - k_\omega ^2 \varepsilon _\omega  \left( {\mathbf{r}} \right)} \right]{\mathcal F}_\omega  \left( {\left. {\mathbf{r}} \right|{\mathbf{n}}} \right) &=  0,   & 
{\mathcal F}_\omega  \left( {\left. {r{\mathbf{m}}} \right|{\mathbf{n}}} \right) & \mathop  \approx \limits_{r \to + \infty } e^{i\left( {k_\omega  {\mathbf{n}}} \right) \cdot \left( {r{\mathbf{m}}} \right)} {\mathcal I}_{\mathbf{n}}  + \frac{{e^{ik_\omega  r} }}{r}{\mathcal S}_\omega  \left( {{\mathbf{m}}\left| {\mathbf{n}} \right.} \right),  \nonumber \\
\left[ {\left( {\nabla _{\mathbf{r}}  \times \frac{1}{{\mu _\omega  \left( {\mathbf{r}} \right)}}\nabla _{\mathbf{r}}  \times } \right) - k_\omega ^2 \varepsilon _\omega  \left( {\mathbf{r}} \right)} \right]{\mathcal G}_\omega  \left( {\left. {\mathbf{r}} \right|{\mathbf{r}}'} \right) & = \delta \left( {{\mathbf{r}} - {\mathbf{r}}'} \right){\mathcal I}, &
{\mathcal G}_\omega  \left( {\left. {r{\mathbf{m}}} \right|{\mathbf{r}}'} \right) & \mathop  \approx \limits_{r \to + \infty } \frac{{e^{ik_\omega  r} }}{r}{\mathcal W}_\omega  \left( {\left. {\mathbf{m}} \right|{\mathbf{r}}'} \right), 
\end{align}
where $\mathbf{n}$ and $\mathbf{m}$ are unit vectors, $\mathop  \approx \limits_{r \to + \infty }$ denotes the leading-order asymptotic behavior, ${\mathcal S}_\omega  \left( {{\mathbf{m}}\left| {\mathbf{n}} \right.} \right)$ is the scattering dyadic \cite{Krist2016}, and ${\mathcal W}_\omega  \left( {{\mathbf{m}}\left| {\mathbf{r}} \right.} \right)$ is the asymptotic amplitude of the Green's function.  The scattering dyadic ${\mathcal S}_\omega  \left( {{\mathbf{m}}\left| {\mathbf{n}} \right.} \right)$ satisfies the orthogonality relations 
\begin{align} \label{TrasvSom}
{\mathbf{m}} \cdot {\mathcal S}_\omega  \left( {{\mathbf{m}}\left| {\mathbf{n}} \right.} \right) &= 0, &
{\mathcal S}_\omega  \left( {{\mathbf{m}}\left| {\mathbf{n}} \right.} \right) \cdot {\mathbf{n}} &= 0 
\end{align}
together with the reiciprocity relation
\begin{equation} \label{RecipSom}
{\cal S}_\omega ^T \left( {{\bf{m}}\left| {\bf{n}} \right.} \right) = {\cal S}_\omega  \left( { - {\bf{n}}\left| { - {\bf{m}}} \right.} \right)
\end{equation}
As a consequence, the modal dyadic satisfies the orthogonality relation ${\mathcal F}_\omega  \left( {\left. {\mathbf r} \right|{\mathbf{n}}} \right) \cdot {\mathbf{n}} = 0$ and accordingly it is convenient to introduce two mutually orthogonal unit vectors ${\mathbf{e}}_{{\mathbf{n}}1}$, ${\mathbf{e}}_{{\mathbf{n}}2}$ orthogonal to ${\mathbf n}$ so that, since  ${\mathcal I}_{\mathbf{n}}  = \sum\nolimits_\lambda  {{\mathbf{e}}_{{\mathbf{n}} \lambda} {\mathbf{e}}_{{\mathbf{n}} \lambda} }$, the field ${\mathbf{F}}_{\omega {\mathbf{n}}\lambda } \left( {\mathbf{r}} \right) = {\mathcal F}_{\omega} \left( {\left. {\mathbf{r}} \right|{\mathbf{n}}} \right) \cdot {\mathbf{e}}_{{\mathbf{n}}\lambda }$ satisfies the boundary value problem
\begin{align}
\left[ {\left( {\nabla _{\mathbf{r}}  \times \frac{1}{{\mu _\omega  \left( {\mathbf{r}} \right)}}\nabla _{\mathbf{r}}  \times } \right) - k_\omega ^2 \varepsilon _\omega  \left( {\mathbf{r}} \right)} \right]{\mathbf{F}}_{\omega {\mathbf{n}}\lambda } \left( {\mathbf{r}} \right) &= 0, &
{\mathbf{F}}_{\omega {\mathbf{n}}\lambda } \left( {r{\mathbf{m}}} \right) & \mathop  \approx \limits_{r \to  + \infty } e^{i\left( {k_\omega  {\mathbf{n}}} \right) \cdot \left( {r{\mathbf{m}}} \right)} {\mathbf{e}}_{{\mathbf{n}}\lambda }  + \frac{{e^{ik_\omega  r} }}{r}{\mathcal S}_\omega  \left( {{\mathbf{m}}\left| {\mathbf{n}} \right.} \right) \cdot {\mathbf{e}}_{{\mathbf{n}}\lambda } 
\end{align}
or, in other words, ${\mathbf{F}}_{\omega {\mathbf{n}}\lambda } \left( {\mathbf{r}} \right)$ is the overall field scattered by the object (scattering mode) when illuminated by the modal plane wave of frequency $\omega$, direction $\mathbf n$ and polarization ${\mathbf{e}}_{{\mathbf{n}}\lambda }$. It is also evident that the modal dyadic admits the decomposition 
${\mathcal F}_\omega  \left( {\left. {\mathbf{r}} \right|{\mathbf{n}}} \right) = \sum\nolimits_\lambda  {{\mathbf{F}}_{\omega {\mathbf{n}}\lambda } \left( {\mathbf{r}} \right){\mathbf{e}}_{{\mathbf{n}}\lambda } }$ which further elucidates its physical meaning. The dyadic ${\mathcal W}_\omega  \left( {{\mathbf{m}}\left| {\mathbf{r}} \right.} \right)$ satisfies the orthogonality relation ${\mathbf{m}} \cdot {\mathcal W}_\omega  \left( {{\mathbf{m}}\left| {\mathbf{r}} \right.} \right) = 0$. Two essential properties of the dyadic Green's function are the reciprocity relation ${\mathcal G}_\omega ^T \left( {\left. {\mathbf{r}} \right|{\mathbf{r}}'} \right) = {\mathcal G}_\omega  \left( {\left. {{\mathbf{r}}'} \right|{\mathbf{r}}} \right)$ and the fundamental integral relation
\begin{equation} \label{Gom_fun}
\frac{{\hbar k_\omega ^3 }}{{\pi \varepsilon _0 }}\int {do_{\mathbf{m}} } {\mathcal W}_\omega ^T \left( {\left. {\mathbf{m}} \right|{\mathbf{r}}} \right) \cdot {\mathcal W}_\omega ^* \left( {\left. {\mathbf{m}} \right|{\mathbf{r}}'} \right) + \sum\limits_\nu  {\int {d^3 {\mathbf{s}}} \,{\mathcal G}_{\omega \nu } \left( {\left. {\mathbf{r}} \right|{\mathbf{s}}} \right) \cdot {\mathcal G}_{\omega \nu }^{T*} \left( {\left. {{\mathbf{r}}'} \right|{\mathbf{s}}} \right)}  = \frac{{\hbar k_\omega ^2 }}{{\pi \varepsilon _0 }}{\mathop{\rm Im}\nolimits} \left[ {{\mathcal G}_\omega  \left( {\left. {\mathbf{r}} \right|{\mathbf{r}}'} \right)} \right],
\end{equation}
where ${\mathcal G}_{\omega e} \left( {\left. {\mathbf{r}} \right|{\mathbf{r}}'} \right)$ and ${\mathcal G}_{\omega m} \left( {\left. {\mathbf{r}} \right|{\mathbf{r}}'} \right)$ are the electric and magnetic dyadic kernels 
\begin{align} \label{GoeGom}
{\mathcal G}_{\omega e} \left( {\left. {\mathbf{r}} \right|{\mathbf{r}}'} \right) =& {\mathcal G}_\omega  \left( {\left. {\mathbf{r}} \right|{\mathbf{r}}'} \right)i\sqrt {\frac{{\hbar k_\omega ^4 }}{{\pi \varepsilon _0 }}{\mathop{\rm Im}\nolimits} \left[ {\varepsilon _\omega  \left( {{\mathbf{r}}'} \right)} \right]} , & {\mathcal G}_{\omega m} \left( {\left. {\mathbf{r}} \right|{\mathbf{r}}'} \right) =& {\mathcal G}_\omega  \left( {\left. {\mathbf{r}} \right|{\mathbf{r}}'} \right)\frac{{ \times \mathord{\buildrel{\lower3pt\hbox{$\scriptscriptstyle\leftarrow$}} 
\over \nabla } _{{\mathbf{r}}'} }}{{ik_\omega  }}\sqrt {\frac{{\hbar k_\omega ^4 }}{{\pi \varepsilon _0 }}{\mathop{\rm Im}\nolimits} \left[ {\frac{{ - 1}}{{\mu _\omega  \left( {{\mathbf{r}}'} \right)}}} \right]}.
\end{align}
Such essential properties have been rederived in Ref.\cite{Ciatt2024} where, in addition, it  has been shown that the asymptotic amplitude of the dyadic Green's function is related to the modal dyadic by the relation
\begin{equation} \label{Womr}
{\mathcal W}_\omega  \left( {{\mathbf{m}}\left| {\mathbf{r}}\right.} \right) = \frac{1}{{4\pi }}{\mathcal F}_\omega ^T \left( {\left. {\mathbf{r}} \right| - {\mathbf{m}}} \right)
\end{equation}
which, together with Eq.(\ref{Gom_fun}), yields the integral relation 
\begin{equation} \label{MLNF_FundIntRel}
\int {do_{\mathbf{n}} } {\mathcal F}_{\omega s} \left( {\left. {\mathbf{r}} \right|{\mathbf{n}}} \right) \cdot {\mathcal F}_{\omega s}^{T*} \left( {\left. {{\mathbf{r}}'} \right|{\mathbf{n}}} \right) + \sum\limits_\nu  {\int {d^3 {\mathbf{s}}} \,{\mathcal G}_{\omega \nu } \left( {\left. {\mathbf{r}} \right|{\mathbf{s}}} \right) \cdot {\mathcal G}_{\omega \nu }^{T*} \left( {\left. {{\mathbf{r}}'} \right|{\mathbf{s}}} \right)}  = \frac{{\hbar k_\omega ^2 }}{{\pi \varepsilon _0 }}{\mathop{\rm Im}\nolimits} \left[ {{\mathcal G}_\omega  \left( {\left. {\mathbf{r}} \right|{\mathbf{r}}'} \right)} \right]
\end{equation}
connecting the modal dyadic and the dyadic Green's function, where ${\mathcal F}_{\omega s} \left( {\left. {\mathbf{r}} \right|{\mathbf{n}}} \right)$ is the scattering dyadic kernel 
\begin{equation} \label{Fos}
{\mathcal F}_{\omega s} \left( {\left. {\mathbf{r}} \right|{\mathbf{n}}} \right) = {\mathcal F}_\omega  \left( {\left. {\mathbf{r}} \right|{\mathbf{n}}} \right)\sqrt {\frac{{\hbar k_\omega ^3 }}{{16\pi ^3 \varepsilon _0 }}} .
\end{equation}

The MLNF describes the field-matter system via the continuous bosonic operators for positive frequencies ($\omega > 0$): the transverse scattering polaritons ${\mathbf{\hat g}}_{\omega s} \left( {\mathbf{n}} \right)$ (${{\mathbf{n}} \cdot {\mathbf{\hat g}}_{\omega s} = 0}$), and the electric ($e$) and magnetic ($m$) medium polaritons ${\mathbf{\hat f}}_{\omega \nu } \left( {\mathbf{r}} \right)$ (with $\nu \in \{e,m\}$), which are strictly confined to the object's volume (${{\mathbf{r}} \in V}$). These operators satisfy the canonical commutation relations
\begin{align} \label{MLNF_Com_Rel}
\left[ {{\mathbf{\hat g}}_{\omega s} \left( {\mathbf{n}} \right),{\mathbf{\hat g}}_{\omega 's}^\dag  \left( {{\mathbf{n}}'} \right)} \right] &= \delta \left( {\omega  - \omega '} \right)\delta \left( {o_{\mathbf{n}}  - o_{{\mathbf{n}}'} } \right){\mathcal I}_{\mathbf{n}}, & \left[ {{\mathbf{\hat f}}_{\omega \nu } \left( {\mathbf{r}} \right),{\mathbf{\hat f}}_{\omega '\nu '}^\dag  \left( {{\mathbf{r}}'} \right)} \right] &= \delta \left( {\omega  - \omega '} \right)\delta _{\nu \nu '} \delta \left( {{\mathbf{r}} - {\mathbf{r}}'} \right){\mathcal I},
\end{align}
with all other commutators vanishing, where ${\mathcal I}_{\mathbf{n}}$ is the dyadic projector onto the plane orthogonal to the direction ${\mathbf{n}} = \sin \theta _{\mathbf{n}} \left( {\cos \varphi _{\mathbf{n}} {\mathbf{u}}_x  + \sin \varphi _{\mathbf{n}} {\mathbf{u}}_y } \right) + \cos \theta _{\mathbf{n}} {\mathbf{u}}_z$,  ${\mathcal I}$  is the dyadic identity and $\delta \left( {o_{\mathbf{n}}  - o_{{\mathbf{n}}'} } \right) = \delta \left( {\theta _{\mathbf{n}}  - \theta '_{\mathbf{n}} } \right)\delta \left( {\varphi _{\mathbf{n}}  - \varphi '_{\mathbf{n}} } \right)/\sin \theta _{\mathbf{n}}$ is the angular delta function. The Hamiltonian operator is
\begin{equation} \label{MLNF_H}
\hat H = \int {d\omega } \;\hbar \omega \left[ {\int {do_{\mathbf{n}} } {\mathbf{\hat g}}_{\omega s}^\dag  \left( {\mathbf{n}} \right) \cdot {\mathbf{\hat g}}_{\omega s} \left( {\mathbf{n}} \right) + \sum\limits_{\nu} {\int {d^3 {\mathbf{r}}\,} {\mathbf{\hat f}}_{\omega \nu }^\dag  \left( {\mathbf{r}} \right) \cdot {\mathbf{\hat f}}_{\omega \nu } \left( {\mathbf{r}} \right)} } \right] 
\end{equation}
where $do_{\mathbf{n}}  = \sin \theta _{\mathbf{n}} d\theta _{\mathbf{n}} d\varphi _{\mathbf{n}}$ is the solid angle element around the direction $\mathbf n$, whereas the electric field and magnetic induction field operators are
\begin{align} \label{MLNF_EB}
 {\mathbf{\hat E}}\left( {\mathbf{r}} \right) &= \int {d\omega } \left[ {{\mathbf{\hat E}}_\omega  \left( {\mathbf{r}} \right) + {\mathbf{\hat E}}_\omega ^\dag  \left( {\mathbf{r}} \right)} \right], &
 {\mathbf{\hat B}}\left( {\mathbf{r}} \right) &= \int {d\omega } \frac{1}{{i\omega }}\left[ {\nabla  \times {\mathbf{\hat E}}_\omega  \left( {\mathbf{r}} \right) - \nabla  \times {\mathbf{\hat E}}_\omega ^\dag  \left( {\mathbf{r}} \right)} \right] 
\end{align}
where
\begin{equation} \label{MLNF_SpeEle}
{\mathbf{\hat E}}_\omega  \left( {\mathbf{r}} \right) = \int {do_{\mathbf{n}} } {\mathcal F}_{\omega s} \left( {\left. {\mathbf{r}} \right|{\mathbf{n}}} \right) \cdot {\mathbf{\hat g}}_{\omega s} \left( {\mathbf{n}} \right) + \sum\limits_\nu  {\int {d^3 {\mathbf{r}}'\,} {\mathcal G}_{\omega \nu } \left( {\left. {\mathbf{r}} \right|{\mathbf{r}}'} \right) \cdot {\mathbf{\hat f}}_{\omega \nu } \left( {{\mathbf{r}}'} \right)}.
\end{equation}
Note that this spectral electric field operator  is consistent with the polariton definitions because, from Eqs.(\ref{Fos}), ${\mathcal F}_{\omega s} \left( {\left. {\mathbf{r}} \right|{\mathbf{n}}} \right) \cdot {\mathbf{n}} = 0$ and, from Eqs.(\ref{GoeGom}), ${\mathcal G}_{\omega \nu } \left( {\left. {\mathbf{r}} \right|{\mathbf{r}}'} \right) = 0$ for ${\mathbf{r}}' \notin V$. As a final remark, we note that the bosonic commutation relations of the polariton operators in Eqs.(\ref{MLNF_Com_Rel}) straightforwardly yield
\begin{align} \label{MLNF_COM_HE}
\left[ {\hat H,{\mathbf{\hat g}}_{\omega s} \left( {\mathbf{n}} \right)} \right] &=  - \hbar \omega {\mathbf{\hat g}}_{\omega s} \left( {\mathbf{n}} \right), &
 \left[ {\hat H,{\mathbf{\hat f}}_{\omega \nu } \left( {\mathbf{r}} \right)} \right] &=  - \hbar \omega {\mathbf{\hat f}}_{\omega \nu } \left( {\mathbf{r}} \right), &
\left[ {\hat H,{\mathbf{\hat E}}_\omega  \left( {\mathbf{r}} \right)} \right] &=  - \hbar \omega {\mathbf{\hat E}}_\omega  \left( {\mathbf{r}} \right), 
\end{align}
whereas the fundamental integral identity in Eq.(\ref{MLNF_FundIntRel}) enables to prove that
\begin{align} \label{MLNF_COM_EE}
 \left[ {{\mathbf{\hat E}}_\omega  \left( {\mathbf{r}} \right),{\mathbf{\hat E}}_{\omega '}^\dag  \left( {{\mathbf{r}}'} \right)} \right] &= \delta \left( {\omega  - \omega '} \right)\frac{{\hbar k_\omega ^2 }}{{\pi \varepsilon _0 }}{\mathop{\rm Im}\nolimits} \left[ {{\mathcal G}_\omega  \left( {{\mathbf{r}}\left| {{\mathbf{r}}'} \right.} \right)} \right] . 
\end{align}

\section{Evaluation of the Maxwell stress dyadic expectation value}

By using the expression for the electric field and the magnetic induction field operators in Eqs.(\ref{MLNF_EB}), the time-averaged expectation value of the Maxwell stress dyadic operator in Eq.(\ref{Maxwell_Tensor}) at a position $\mathbf{r}$ in the vacuum region outside the objects turns out to be
\begin{equation} \label{TA}
\langle\langle {\hat {\mathcal T}\left( {\mathbf{r}} \right)} \rangle\rangle  = \varepsilon _0 \int {d\omega } \int {d\omega '} \;\left\{ {{\mathcal A}_{\omega ,\omega '} \left( {{\mathbf{r}}, {\mathbf{r}}'} \right) - \frac{1}{2}\text{Tr}\left[ {{\mathcal A}_{\omega ,\omega '} \left( {{\mathbf{r}}, {\mathbf{r}}'} \right)} \right]{\mathcal I}} \right\}_{{\mathbf{r}}' \to {\mathbf{r}}},
\end{equation}
where the standard prescription ${\mathbf{r}}' \to {\mathbf{r}}$ has been used, while the spectral correlation dyadic ${\mathcal A}_{\omega ,\omega '}\left( {{\mathbf{r}}, {\mathbf{r}}'} \right)$ is given by
\begin{eqnarray} \label{A_omom'}
&& {\mathcal A}_{\omega ,\omega '} {\left( {{\mathbf{r}}, {\mathbf{r}}'} \right)} = \left[ {\langle\langle {{\mathbf{\hat E}}_\omega  \left( {\mathbf{r}} \right){\mathbf{\hat E}}_{\omega '} \left( {{\mathbf{r}}'} \right)} \rangle\rangle  + \langle\langle {{\mathbf{\hat E}}_\omega ^\dag  \left( {\mathbf{r}} \right){\mathbf{\hat E}}_{\omega '}^\dag  \left( {{\mathbf{r}}'} \right)} \rangle\rangle  + \langle\langle {{\mathbf{\hat E}}_\omega ^\dag  \left( {\mathbf{r}} \right){\mathbf{\hat E}}_{\omega '} \left( {{\mathbf{r}}'} \right)} \rangle\rangle  + \langle\langle {{\mathbf{\hat E}}_\omega  \left( {\mathbf{r}} \right){\mathbf{\hat E}}_{\omega '}^\dag  \left( {{\mathbf{r}}'} \right)} \rangle\rangle } \right] \nonumber  \\ 
  &+& \frac{1}{{k_\omega  k_{\omega '} }}\nabla _{\mathbf{r}}  \times \left[ {\langle\langle {{\mathbf{\hat E}}_\omega  \left( {\mathbf{r}} \right){\mathbf{\hat E}}_{\omega '} \left( {{\mathbf{r}}'} \right)} \rangle\rangle  + \langle\langle {{\mathbf{\hat E}}_\omega ^\dag  \left( {\mathbf{r}} \right){\mathbf{\hat E}}_{\omega '}^\dag  \left( {{\mathbf{r}}'} \right)} \rangle\rangle  - \langle\langle {{\mathbf{\hat E}}_\omega ^\dag  \left( {\mathbf{r}} \right){\mathbf{\hat E}}_{\omega '} \left( {{\mathbf{r}}'} \right)} \rangle\rangle  - \langle\langle {{\mathbf{\hat E}}_\omega  \left( {\mathbf{r}} \right){\mathbf{\hat E}}_{\omega '}^\dag  \left( {{\mathbf{r}}'} \right)} \rangle\rangle } \right] \times \mathord{\buildrel{\lower3pt\hbox{$\scriptscriptstyle\leftarrow$}} 
\over \nabla } _{{\mathbf{r}}'}. 
\end{eqnarray}
It is here convenient to introduce the time-averaged total density operator
\begin{equation}
\overline {\hat \rho }  = \left[ {\mathop {\lim }\limits_{T \to  + \infty } \frac{1}{T}\int_{ - T/2}^{T/2} {dt} \hat \rho _s \left( t \right)} \right]\hat \rho _{em} 
\end{equation}
such that for any operator $\hat O$ the relation $\langle\langle {\hat O} \rangle\rangle  = {\rm Tr} ( {\overline {\hat \rho } \hat O} )$ strictly holds. The underlying motivation for such a definition is that, as can be readily demonstrated from the Liouville-von Neumann equation, this time-averaged operator commutes with the Hamiltonian $[ {\hat H,\overline {\hat \rho } } ] = 0 $, which is a crucial property involved in the subsequent evaluation of the four dyadic correlators appearing in Eq.(\ref{A_omom'}). 

Let us first consider the evaluation of the anomalous correlator $\langle\langle \hat{\mathbf{E}}_\omega(\mathbf{r}) \hat{\mathbf{E}}_{\omega'}(\mathbf{r}') \rangle\rangle$. By expressing its left electric field operator through the relation ${\mathbf{\hat E}}_\omega  \left( {\mathbf{r}} \right) =  - \frac{1}{{\hbar \omega }}[ {\hat H,{\mathbf{\hat E}}_\omega  ( {\mathbf{r}} )} ]$ (see the third of Eqs.(\ref{MLNF_COM_HE})), invoking the stationarity property $[\hat{H}, \overline{\hat{\rho}}] = 0$, and exploiting the cyclic property of the trace, the Hamiltonian commutator is effectively transferred to the right electric field operator, yielding
\begin{equation}
\langle\langle \hat{\mathbf{E}}_\omega(\mathbf{r}) \hat{\mathbf{E}}_{\omega'}(\mathbf{r}') \rangle\rangle = -\frac{1}{\hbar\omega} \text{Tr} \left\{ \overline{\hat{\rho}} [\hat{H}, \hat{\mathbf{E}}_\omega(\mathbf{r})] \hat{\mathbf{E}}_{\omega'}(\mathbf{r}') \right\} = \frac{1}{\hbar\omega} \text{Tr} \left\{ \overline{\hat{\rho}} \hat{\mathbf{E}}_\omega(\mathbf{r}) [\hat{H}, \hat{\mathbf{E}}_{\omega'}(\mathbf{r}')] \right\} = -\frac{\omega'}{\omega} \langle\langle \hat{\mathbf{E}}_\omega(\mathbf{r}) \hat{\mathbf{E}}_{\omega'}(\mathbf{r}') \rangle\rangle.
\end{equation}
Since frequencies are strictly positive ($\omega, \omega' > 0$), this relation necessarily implies 
\begin{equation} \label{CEE}
\langle\langle \hat{\mathbf{E}}_\omega(\mathbf{r}) \hat{\mathbf{E}}_{\omega'}(\mathbf{r}') \rangle\rangle = 0.
\end{equation}
By taking the Hermitian conjugate of this result, it trivially follows that 
\begin{equation} \label{CE+E+}
\langle\langle \hat{\mathbf{E}}_\omega^\dagger(\mathbf{r}) \hat{\mathbf{E}}_{\omega'}^\dagger(\mathbf{r}') \rangle\rangle = 0
\end{equation} 
as well. Turning to the normally ordered correlator ${\langle\langle {{\mathbf{\hat E}}_\omega ^\dag \left( {\mathbf{r}} \right){\mathbf{\hat E}}_{\omega '} \left( {{\mathbf{r}}'} \right)} \rangle\rangle }$, substituting the expression for the spectral electric field operator from Eq.(\ref{MLNF_SpeEle}) and exploiting the statistical independence between the scattering and medium-assisted sectors yields
\begin{eqnarray}
 \langle\langle {{\mathbf{\hat E}}_\omega ^\dag  \left( {\mathbf{r}} \right){\mathbf{\hat E}}_{\omega '} \left( {{\mathbf{r}}'} \right)} \rangle\rangle  &=& \int {do_{\mathbf{n}} } \int {do_{{\mathbf{n}}'} } {\mathcal F}_{\omega s}^* \left( {\left. {\mathbf{r}} \right|{\mathbf{n}}} \right) \cdot \langle\langle {{\mathbf{\hat g}}_{\omega s}^\dag  \left( {\mathbf{n}} \right){\mathbf{\hat g}}_{\omega 's} \left( {{\mathbf{n}}'} \right)} \rangle\rangle  \cdot {\mathcal F}_{\omega 's}^T \left( {\left. {{\mathbf{r}}'} \right|{\mathbf{n}}'} \right) \nonumber \\ 
  &+& \int {d^3 {\mathbf{s}}\,} \sum\limits_\nu  {} \int {d^3 {\mathbf{s}}'\,} \sum\limits_{\nu '} {} {\mathcal G}_{\omega \nu }^* \left( {\left. {\mathbf{r}} \right|{\mathbf{s}}} \right) \cdot \langle\langle {{\mathbf{\hat f}}_{\omega \nu }^\dag  \left( {\mathbf{s}} \right){\mathbf{\hat f}}_{\omega '\nu '} \left( {{\mathbf{s}}'} \right)} \rangle\rangle  \cdot {\mathcal G}_{\omega '\nu '}^T \left( {\left. {{\mathbf{r}}'} \right|{\mathbf{s}}'} \right) \nonumber  \\ 
  &+& \left\{ {\int {d^3 {\mathbf{s}}\,} \sum\limits_\nu  {{\mathcal G}_{\omega \nu } \left( {\left. {\mathbf{r}} \right|{\mathbf{s}}} \right) \cdot \langle\langle {{\mathbf{\hat f}}_{\omega \nu } \left( {\mathbf{s}} \right)} \rangle\rangle } } \right\}^* \left\{ {\int {do_{\mathbf{n}} } {\mathcal F}_{\omega 's} \left( {\left. {{\mathbf{r}}'} \right|{\mathbf{n}}} \right) \cdot \langle\langle {{\mathbf{\hat g}}_{\omega 's} \left( {\mathbf{n}} \right)} \rangle\rangle } \right\} \nonumber  \\ 
  &+& \left\{ {\int {do_{\mathbf{n}} } {\mathcal F}_{\omega s} \left( {\left. {\mathbf{r}} \right|{\mathbf{n}}} \right) \cdot \langle\langle {{\mathbf{\hat g}}_{\omega s} \left( {\mathbf{n}} \right)} \rangle\rangle } \right\}^* \left\{ {\int {d^3 {\mathbf{s}}\,} \sum\limits_\nu  {{\mathcal G}_{\omega '\nu } \left( {\left. {{\mathbf{r}}'} \right|{\mathbf{s}}} \right) \cdot  \langle\langle {{\mathbf{\hat f}}_{\omega '\nu } \left( {\mathbf{s}} \right)} \rangle\rangle } } \right\}, 
\end{eqnarray}
consisting of four distinct terms: the pure scattering contribution, the pure material fluctuation term, and the cross-correlations between the two sectors. By applying the thermal averages of Eqs.(\ref{thermal_averages}) (which, due to the stationarity of the medium bath, correspond identically to their time-averaged counterparts, e.g., $\langle\langle \mathbf{\hat{f}}^\dagger \mathbf{\hat{f}} \rangle\rangle = \langle \mathbf{\hat{f}}^\dagger \mathbf{\hat{f}} \rangle$), the cross-terms identically vanish and the Bose-Einstein distribution shows up in the material fluctuation term, so that we obtain
\begin{eqnarray} \label{E+E}
 \langle\langle {{\mathbf{\hat E}}_\omega ^\dag  \left( {\mathbf{r}} \right){\mathbf{\hat E}}_{\omega '} \left( {{\mathbf{r}}'} \right)} \rangle\rangle  &=& \left[ {\int {do_{\mathbf{n}} } \int {do_{{\mathbf{n}}'} } {\mathcal F}_{\omega s} \left( {\left. {\mathbf{r}} \right|{\mathbf{n}}} \right) \cdot \langle\langle {{\mathbf{\hat g}}_{\omega s}^\dag  \left( {\mathbf{n}} \right){\mathbf{\hat g}}_{\omega 's} \left( {{\mathbf{n}}'} \right)} \rangle\rangle ^* \cdot {\mathcal F}_{\omega 's}^{T*} \left( {\left. {{\mathbf{r}}'} \right|{\mathbf{n}}'} \right)} \right. \nonumber \\ 
 &+& \left. {  \delta \left( {\omega  - \omega '} \right)n_\omega  \left( {\beta _{em} } \right)\int {d^3 {\mathbf{s}}\,} \sum\limits_\nu  {{\mathcal G}_{\omega \nu } \left( {\left. {\mathbf{r}} \right|{\mathbf{s}}} \right) \cdot {\mathcal G}_{\omega \nu }^{T*} \left( {\left. {{\mathbf{r}}'} \right|{\mathbf{s}}} \right)} } \right]^*. 
\end{eqnarray}
The scattering correlator is diagonal with respect to frequency, i.e.,
\begin{equation} \label{diag_spec_corr}
\left\langle\left\langle {{\mathbf{\hat g}}_{\omega s}^\dag  \left( {\mathbf{n}} \right){\mathbf{\hat g}}_{\omega 's} \left( {{\mathbf{n}}'} \right)} \right\rangle\right\rangle  = \delta \left( {\omega  - \omega '} \right)\int {d\omega '} \left\langle\left\langle {{\mathbf{\hat g}}_{\omega s}^\dag  \left( {\mathbf{n}} \right){\mathbf{\hat g}}_{\omega 's} \left( {{\mathbf{n}}'} \right)} \right\rangle\right\rangle 
\end{equation}
as can be demonstrated by following a procedure analogous to the one used to prove that the anomalous dyadic correlator  vanishes. Indeed, by expressing the adjoint scattering operator through the commutator ${\mathbf{\hat g}}_{\omega s}^\dag  \left( {\mathbf{n}} \right) = \frac{1}{{\hbar \omega }} [ {\hat H,{\mathbf{\hat g}}_{\omega s}^\dag  \left( {\mathbf{n}} \right)}  ]$ (see the first of Eqs.(\ref{MLNF_COM_HE})) and exploiting the stationarity of the time-averaged density operator, $[\hat{H}, \overline{\hat{\rho}}] = 0$, we can write
\begin{equation}
\langle\langle \hat{\mathbf{g}}_{\omega s}^\dagger(\mathbf{n}) \hat{\mathbf{g}}_{\omega' s}(\mathbf{n}') \rangle\rangle = \frac{1}{\hbar\omega} \text{Tr} \left\{ \overline{\hat{\rho}} [\hat{H}, \hat{\mathbf{g}}_{\omega s}^\dagger(\mathbf{n})] \hat{\mathbf{g}}_{\omega' s}(\mathbf{n}') \right\} = -\frac{1}{\hbar\omega} \text{Tr} \left\{ \overline{\hat{\rho}} \hat{\mathbf{g}}_{\omega s}^\dagger(\mathbf{n}) [\hat{H}, \hat{\mathbf{g}}_{\omega' s}(\mathbf{n}')] \right\} = \frac{\omega'}{\omega} \langle\langle \hat{\mathbf{g}}_{\omega s}^\dagger(\mathbf{n}) \hat{\mathbf{g}}_{\omega' s}(\mathbf{n}') \rangle\rangle,
\end{equation}
where the cyclic property of the trace has been used to transfer the action of the Hamiltonian. This result leads to the condition $(\omega - \omega') \langle\langle \hat{\mathbf{g}}_{\omega s}^\dagger(\mathbf{n}) \hat{\mathbf{g}}_{\omega' s}(\mathbf{n}') \rangle\rangle = 0$, whose solution in the sense of distributions is provided by Eq.(\ref{diag_spec_corr}). Inserting Eq.(\ref{diag_spec_corr}) into Eq.(\ref{E+E}), we obtain
\begin{eqnarray} \label{CE+E}
 \langle\langle {{\mathbf{\hat E}}_\omega ^\dag  \left( {\mathbf{r}} \right){\mathbf{\hat E}}_{\omega '} \left( {{\mathbf{r}}'} \right)} \rangle\rangle  &=& \delta \left( {\omega  - \omega '} \right)\left[ {\int {do_{\mathbf{n}} } \int {do_{{\mathbf{n}}'} } {\mathcal F}_{\omega s} \left( {\left. {\mathbf{r}} \right|{\mathbf{n}}} \right) \cdot \int {d\omega '} \langle\langle {{\mathbf{\hat g}}_{\omega s}^\dag  \left( {\mathbf{n}} \right){\mathbf{\hat g}}_{\omega 's} \left( {{\mathbf{n}}'} \right)} \rangle\rangle ^* \cdot {\mathcal F}_{\omega s}^{T*} \left( {\left. {{\mathbf{r}}'} \right|{\mathbf{n}}'} \right)} \right. \nonumber \\ 
&+& \left. {  n_\omega  \left( {\beta _{em} } \right)\int {d^3 {\mathbf{s}}\,} \sum\limits_\nu  {{\mathcal G}_{\omega \nu } \left( {\left. {\mathbf{r}} \right|{\mathbf{s}}} \right) \cdot {\mathcal G}_{\omega \nu }^{T*} \left( {\left. {{\mathbf{r}}'} \right|{\mathbf{s}}} \right)} } \right]^*.
\end{eqnarray}
Finally, the anti-normally ordered dyadic correlator $\langle\langle {{\mathbf{\hat E}}_\omega  \left( {\mathbf{r}} \right){\mathbf{\hat E}}_{\omega '}^\dag  \left( {{\mathbf{r}}'} \right)} \rangle\rangle$ can be evaluated by expressing the dyadic product of the field operators as
\begin{equation}
{\mathbf{\hat E}}_\omega \left( {\mathbf{r}} \right){\mathbf{\hat E}}_{\omega '}^\dag \left( {{\mathbf{r}}'} \right) = \left[ {{\mathbf{\hat E}}_\omega \left( {\mathbf{r}} \right),{\mathbf{\hat E}}_{\omega '}^\dag \left( {{\mathbf{r}}'} \right)} \right] + \left\{ {{\mathbf{\hat E}}_{\omega '}^\dag \left( {{\mathbf{r}}'} \right){\mathbf{\hat E}}_\omega \left( {\mathbf{r}} \right)} \right\}^T,
\end{equation}
which holds by definition of the dyadic commutator. By evaluating the commutator through Eq.(\ref{MLNF_COM_EE}) and applying the general property of the operator trace $\left[ {\text{Tr}\left( {\hat \rho {\mathbf{\hat A\hat B}}} \right)} \right]^T = \left[ {\text{Tr}\left( {\hat \rho {\mathbf{\hat B}}^\dag {\mathbf{\hat A}}^\dag } \right)} \right]^*$, we obtain
\begin{align}  \label{CEE+}
\langle\langle {\mathbf{\hat E}}_\omega \left( {\mathbf{r}} \right){\mathbf{\hat E}}_{\omega '}^\dag \left( {{\mathbf{r}}'} \right) \rangle\rangle &= \text{Tr}\left\{ {\overline {\hat \rho } \left\{ {\delta \left( {\omega - \omega '} \right)\frac{{\hbar k_\omega ^2 }}{{\pi \varepsilon _0 }}{\mathop{\rm Im}\nolimits} \left[ {{\mathcal G}_\omega \left( {{\mathbf{r}}\left| {{\mathbf{r}}'} \right.} \right)} \right] + \left[ {{\mathbf{\hat E}}_{\omega '}^\dag \left( {{\mathbf{r}}'} \right){\mathbf{\hat E}}_\omega \left( {\mathbf{r}} \right)} \right]^T } \right\}} \right\} \nonumber \\
&= \delta \left( {\omega - \omega '} \right)\frac{{\hbar k_\omega ^2 }}{{\pi \varepsilon _0 }}{\mathop{\rm Im}\nolimits} \left[ {{\mathcal G}_\omega \left( {{\mathbf{r}}\left| {{\mathbf{r}}'} \right.} \right)} \right] + \left\{ {\text{Tr}\left[ {\overline {\hat \rho } {\mathbf{\hat E}}_{\omega '}^\dag \left( {{\mathbf{r}}'} \right){\mathbf{\hat E}}_\omega \left( {\mathbf{r}} \right)} \right]} \right\}^T \nonumber \\
&= \delta \left( {\omega - \omega '} \right)\frac{{\hbar k_\omega ^2 }}{{\pi \varepsilon _0 }}{\mathop{\rm Im}\nolimits} \left[ {{\mathcal G}_\omega \left( {{\mathbf{r}}\left| {{\mathbf{r}}'} \right.} \right)} \right] + \langle\langle {\mathbf{\hat E}}_\omega ^\dag \left( {\mathbf{r}} \right){\mathbf{\hat E}}_{\omega '} \left( {{\mathbf{r}}'} \right) \rangle\rangle^*.
\end{align}
which provides a fundamental relation connecting the anti-normally ordered dyadic correlator directly to the normally ordered one. Substituting Eqs.(\ref{CEE}), (\ref{CE+E+}), (\ref{CE+E}), and (\ref{CEE+}) into Eq.(\ref{A_omom'}), we obtain
\begin{equation} \label{Afin}
{\mathcal A}_{\omega ,\omega '} \left( {{\mathbf{r}}\left| {{\mathbf{r}}'} \right.} \right) = \delta \left( {\omega  - \omega '} \right)\left[ {{\mathcal C}_\omega  \left( {{\mathbf{r}}\left| {{\mathbf{r}}'} \right.} \right) - \frac{1}{{k_\omega ^2 }}\nabla _{\mathbf{r}}  \times {\mathcal C}_\omega  \left( {{\mathbf{r}}\left| {{\mathbf{r}}'} \right.} \right) \times \mathord{\buildrel{\lower3pt\hbox{$\scriptscriptstyle\leftarrow$}} 
\over \nabla } _{{\mathbf{r}}'} } \right]
\end{equation}
where 
\begin{eqnarray}
 {\mathcal C}_\omega  \left( {{\mathbf{r}}\left| {{\mathbf{r}}'} \right.} \right) &=& \frac{{\hbar k_\omega ^2 }}{{\pi \varepsilon _0 }}{\mathop{\rm Im}\nolimits} \left[ {{\mathcal G}_\omega  \left( {{\mathbf{r}}\left| {{\mathbf{r}}'} \right.} \right)} \right] + 2{\mathop{\rm Re}\nolimits} \int {do_{\mathbf{n}} } \int {do_{{\mathbf{n}}'} } {\mathcal F}_{\omega s} \left( {\left. {\mathbf{r}} \right|{\mathbf{n}}} \right) \cdot \int {d\omega '} \left\langle\left\langle {{\mathbf{\hat g}}_{\omega s}^\dag  \left( {\mathbf{n}} \right){\mathbf{\hat g}}_{\omega 's} \left( {{\mathbf{n}}'} \right)} \right\rangle\right\rangle ^* \cdot {\mathcal F}_{\omega s}^{T*} \left( {\left. {{\mathbf{r}}'} \right|{\mathbf{n}}'} \right)  \nonumber \\ 
  &+& 2n_\omega  \left( {\beta _{em} } \right){\mathop{\rm Re}\nolimits} \int {d^3 {\mathbf{s}}\,} \sum\limits_\nu  {{\mathcal G}_{\omega \nu } \left( {\left. {\mathbf{r}} \right|{\mathbf{s}}} \right) \cdot {\mathcal G}_{\omega \nu }^{T*} \left( {\left. {{\mathbf{r}}'} \right|{\mathbf{s}}} \right)},
\end{eqnarray}
which exactly coincides with Eq.(\ref{Com}). Furthermore, by substituting Eq.(\ref{Afin}) into Eq.(\ref{TA}), we directly obtain Eq.(\ref{AvT}).

\section{Evaluation of the force $\mathbf{F}$}
To calculate the net optomechanical force $\mathbf{F}$ acting on the object, it is mathematically convenient to separate the total momentum flux into its distinct physical contributions stemming from the scattering and the medium-assisted sectors, such that $\mathbf{F} = \mathbf{F}_s + \mathbf{F}_{em}$. Specifically, by substituting Eq.(\ref{Com_Squeezed}) into Eq.(\ref{AvT}) and subsequently into Eq.(\ref{Force}), these two force components are given by
\begin{align} \label{Forces_s_em}
\mathbf{F}_s &= \varepsilon_0 \int d\omega \left\{ \lim_{r \to \infty} r^2 \int do_{\mathbf{r}} \, \mathbf{u}_r \cdot \left[ \mathcal{M}_{\omega s}(\mathbf{r}|\mathbf{r}') - \frac{1}{2} \text{tr} \left[ \mathcal{M}_{\omega s}(\mathbf{r}|\mathbf{r}') \right] \mathcal{I} \right]_{\mathbf{r}' \to \mathbf{r}} \right\}, \nonumber \\
\mathbf{F}_{em} &= \varepsilon_0 \int d\omega \left\{ \lim_{r \to \infty} r^2 \int do_{\mathbf{r}} \, \mathbf{u}_r \cdot \left[ \mathcal{M}_{\omega em}(\mathbf{r}|\mathbf{r}') - \frac{1}{2} \text{tr} \left[ \mathcal{M}_{\omega em}(\mathbf{r}|\mathbf{r}') \right] \mathcal{I} \right]_{\mathbf{r}' \to \mathbf{r}} \right\},
\end{align}
where the spectral density dyadics $\mathcal{M}_{\omega s} $ and $\mathcal{M}_{\omega em}$ are defined as
\begin{align} \label{Dyadics_M}
\mathcal{M}_{\omega s}(\mathbf{r}|\mathbf{r}') &= \text{Re} \int do_{\mathbf{n}} \, W_{\omega s}(\mathbf{n}) \left[ \mathcal{F}_{\omega s}(\mathbf{r}|\mathbf{n}) \cdot \mathcal{F}_{\omega s}^{T*}(\mathbf{r}'|\mathbf{n}) - \frac{1}{k_\omega^2} \nabla_{\mathbf{r}} \times \mathcal{F}_{\omega s}(\mathbf{r}|\mathbf{n}) \cdot \mathcal{F}_{\omega s}^{T*}(\mathbf{r}'|\mathbf{n}) \times \overleftarrow{\nabla}_{\mathbf{r}'} \right], \nonumber \\
\mathcal{M}_{\omega em}(\mathbf{r}|\mathbf{r}') &= W_{\omega em} \, \text{Re} \int d^3\mathbf{s} \sum_\nu \left[ \mathcal{G}_{\omega \nu}(\mathbf{r}|\mathbf{s}) \cdot \mathcal{G}_{\omega \nu}^{T*}(\mathbf{r}'|\mathbf{s}) - \frac{1}{k_\omega^2} \nabla_{\mathbf{r}} \times \mathcal{G}_{\omega \nu}(\mathbf{r}|\mathbf{s}) \cdot \mathcal{G}_{\omega \nu}^{T*}(\mathbf{r}'|\mathbf{s}) \times \overleftarrow{\nabla}_{\mathbf{r}'} \right],
\end{align}
and the corresponding quantum-statistical weights are
\begin{align} \label{Weights_W}
W_{\omega s}(\mathbf{n}) &= 1 + 2 \sinh^2[r_\omega(\mathbf{n})], & 
W_{\omega em} &= 1 + 2 n_\omega(\beta_{em}).
\end{align}
From a physical standpoint, these two force components describe fundamentally distinct interaction mechanisms. The scattering term $\mathbf{F}_s$ represents the active macroscopic push exerted by the external environment. It originates from the momentum transferred to the object during the scattering of the externally structured quantum fluctuations, specifically the injected squeezed vacuum characterized by the angularly dependent driving weight $W_{\omega s}(\mathbf{n})$. Conversely, the medium-assisted term $\mathbf{F}_{em}$ captures the thermodynamic radiation reaction, or thermal recoil. It emerges from the internal fluctuating currents within the lossy volume that continuously emit photons into the surrounding space to maintain local thermodynamic equilibrium. If the object possesses geometric or material asymmetries, this thermal emission is inherently anisotropic, thereby imparting a net mechanical recoil to the body governed by the isotropic thermal weight $W_{\omega em}$.

\subsection{Contribution from the scattering sector}
To evaluate the scattering contribution to the force, $\mathbf{F}_s$, we utilize the asymptotic behavior of the modal dyadic ${\cal F}_\omega (\mathbf{r}|\mathbf{n})$ as defined in Eq.(\ref{GS}), together with the definition of the scattering dyadic kernel ${\cal F}_{\omega s}(\mathbf{r}|\mathbf{n})$ in Eq.(\ref{Fos}). By substituting these asymptotic expressions into the first term of $\mathcal{M}_{\omega s}$ in Eq.(\ref{Dyadics_M}), we obtain
\begin{multline} \label{Asymptotic_F_F_Full}
\text{Re} \int do_{\mathbf{n}} W_\omega(\mathbf{n}) \mathcal{F}_{\omega s}(\mathbf{r}|\mathbf{n}) \cdot \mathcal{F}_{\omega s}^{T*}(\mathbf{r}'|\mathbf{n}) \mathop  \approx \limits_{r,r' \to \infty } \frac{\hbar k_\omega^3}{16\pi^3 \varepsilon_0} \text{Re} \int do_{\mathbf{n}} W_\omega(\mathbf{n}) \left[ e^{i (k_\omega \mathbf{n}) \cdot (\mathbf{r} - \mathbf{r}')} \mathcal{I}_{\mathbf{n}} \right. \\
\left. + \frac{e^{i k_\omega (r - r')}}{rr'} \mathcal{S}_\omega(\mathbf{u}_r | \mathbf{n}) \cdot \mathcal{S}_\omega^{T*}(\mathbf{u}_{r'} | \mathbf{n}) + \frac{e^{-ik_\omega r'}}{r'} e^{i (k_\omega \mathbf{n}) \cdot \mathbf{r}} \mathcal{S}_\omega^{T*}(\mathbf{u}_{r'} | \mathbf{n}) + \frac{e^{ik_\omega r}}{r} e^{-i (k_\omega \mathbf{n}) \cdot \mathbf{r}'} \mathcal{S}_\omega(\mathbf{u}_r | \mathbf{n}) \right].
\end{multline}
The angular integrals involving the rapidly oscillating phase factors $e^{\pm i k_\omega \mathbf{n} \cdot \mathbf{r}}$ and $e^{\mp i k_\omega \mathbf{n} \cdot \mathbf{r}'}$ can be evaluated by invoking the Jones lemma (see Appendix XII of Ref.\cite{Bornn2019}), which provides the asymptotic identity
\begin{equation} \label{Jones_Lemma}
e^{ik_\omega  {\bf{n}} \cdot {\bf{r}}} \mathop  \approx \limits_{r \to \infty } \frac{{2\pi }}{{ik_\omega  }}\left[ {\frac{{e^{ik_\omega  r} }}{r}\delta \left( {o_{\bf{n}}  - o_{{\bf{u}}_r } } \right) - \frac{{e^{ - ik_\omega  r} }}{r}\delta \left( {o_{\bf{n}}  - o_{ - {\bf{u}}_r } } \right)} \right].
\end{equation}
Equation (\ref{Jones_Lemma}) allows for the evaluation of the angular integrals in the last two terms of Eq.(\ref{Asymptotic_F_F_Full}). Moreover, the second (magnetic) term of $\mathcal{M}_{\omega s}$ in Eq.(\ref{Dyadics_M}) can be calculated by noting that in the far-field limit the action of the del operator $\nabla_{\mathbf{r}} $ effectively replaces $\nabla_{\mathbf{r}} \times$ with $(i k_\omega \mathbf{u}_r) \times$ (and similarly $\times \overleftarrow{\nabla}_{\mathbf{r}'}$ with $\times (- i k_\omega \mathbf{u}_{r'}) $). Ultimately, the full spectral density dyadic for the scattering sector $\mathcal{M}_{\omega s}(\mathbf{r}|\mathbf{r}')$ is found to be
\begin{align} \label{Mos_Asymptotic_Full}
\mathcal{M}_{\omega s}(\mathbf{r}|\mathbf{r}') &\mathop  \approx \limits_{r,r' \to \infty } \frac{\hbar k_\omega^3}{16\pi^3 \varepsilon_0} \text{Re} \Biggl\{ \int {do_{\bf{n}} } W_\omega  \left( {\bf{n}} \right)e^{i\left( {k_\omega  {\bf{n}}} \right) \cdot \left( {{\bf{r}} - {\bf{r}}'} \right)} 2 {\cal I}_{\bf{n}}  + \frac{e^{ik_\omega (r - r')}}{rr'} \left[ {{\cal Q}_\omega \left( {{\bf{u}}_r \left| {{\bf{u}}_{r'} } \right.} \right)} - \mathbf{u}_r \times {{\cal Q}_\omega \left( {{\bf{u}}_r \left| {{\bf{u}}_{r'} } \right.} \right)} \times \mathbf{u}_{r'} \right] \nonumber \\
&\quad + \frac{2\pi i}{k_\omega rr'} \Biggl\{ e^{-ik_\omega (r+r')} W_\omega(-\mathbf{u}_r) \left[ \mathcal{S}_\omega^{T*}(\mathbf{u}_{r'} | -\mathbf{u}_r) + \mathbf{u}_r \times \mathcal{S}_\omega^{T*}(\mathbf{u}_{r'} | -\mathbf{u}_r) \times \mathbf{u}_{r'} \right] \nonumber \\
&\quad - e^{ik_\omega (r+r')} W_\omega(-\mathbf{u}_{r'}) \left( \mathcal{S}_\omega(\mathbf{u}_r | -\mathbf{u}_{r'}) + \mathbf{u}_r \times \mathcal{S}_\omega(\mathbf{u}_r | -\mathbf{u}_{r'}) \times \mathbf{u}_{r'} \right) \Biggr\} \Biggr\},
\end{align}
where we have introduced the auxiliary dyadic
\begin{equation}
{{\cal Q}_\omega \left( {{\bf{u}}_r \left| {{\bf{u}}_{r'} } \right.} \right)} = \int do_{\mathbf{n}} W_\omega(\mathbf{n}) \mathcal{S}_\omega(\mathbf{u}_r | \mathbf{n}) \cdot \mathcal{S}_\omega^{T*}(\mathbf{u}_{r'} | \mathbf{n}) + \frac{2\pi i}{k_\omega} \left[ W_\omega(\mathbf{u}_{r'}) \mathcal{S}_\omega(\mathbf{u}_r | \mathbf{u}_{r'}) - W_\omega(\mathbf{u}_r) \mathcal{S}_\omega^{T*}(\mathbf{u}_{r'} | \mathbf{u}_r) \right].
\end{equation}
By evaluating the limit $\mathbf{r}' \to \mathbf{r}$ in the spectral density dyadic $\mathcal{M}_{\omega s}$, we can project the momentum flux along the radial direction. Utilizing the left orthogonality of the scattering dyadic, ${\bf{u}}_r  \cdot {\cal S}_\omega  \left( {\left. {{\bf{u}}_r } \right|{\bf{n}}} \right) = 0$  (see the first of Eqs.(\ref{TrasvSom})), the radial projection simplifies to
\begin{equation} \label{urM}
\mathbf{u}_r \cdot \left[ \mathcal{M}_{\omega s}(\mathbf{r}|\mathbf{r}') \right]_{\mathbf{r}' \to \mathbf{r}} \mathop \approx \limits_{r\to \infty } \frac{\hbar k_\omega^3}{16\pi^3 \varepsilon_0} \mathbf{u}_r \cdot  \int do_{\mathbf{n}} W_\omega(\mathbf{n})  2 \mathcal{I}_{\mathbf{n}}
\end{equation}
Similarly, the trace of the dyadic $\mathcal{M}_{\omega s}$ evaluated at $\mathbf{r}' = \mathbf{r}$ is obtained by applying the trace identity $\text{tr}(\mathbf{V} \times \mathcal{A} \times \mathbf{V}) = \mathbf{V} \cdot \mathcal{A} \cdot \mathbf{V} - \mathbf{V} \cdot \mathbf{V}(\text{tr}\mathcal{A})$, which exactly cancels the highly oscillating terms $e^{\pm i 2k_\omega r}$, yielding
\begin{equation} \label{trM}
{\rm tr}\left[ {{\cal M}_{\omega s} \left( {{\bf{r}}\left| {{\bf{r}}'} \right.} \right)} \right]_{{\bf{r}}' \to {\bf{r}}} \mathop \approx \limits_{r\to \infty } \frac{{\hbar k_\omega ^3 }}{{16\pi ^3 \varepsilon _0 }}\left[ {\int {do_{\bf{n}} } 4W_\omega  \left( {\bf{n}} \right) + \frac{2}{{r^2 }} {\rm tr} {\cal Q}_\omega  \left( {{\bf{u}}_r \left| {{\bf{u}}_r } \right.} \right)} \right],
\end{equation}
where we have used the reality of ${\cal Q}_\omega \left( {{\bf{u}}_r \left| {{\bf{u}}_r } \right.} \right)$ and the fact that the trace of the transverse identity is $\text{tr}(\mathcal{I}_{\mathbf{n}}) = 2$. Substituting Eqs.(\ref{urM}) and (\ref{trM}) into the definition of the scattering force $\mathbf{F}_s$ in Eq.(\ref{Forces_s_em}), after some algebra we obtain
\begin{equation}
{\bf{F}}_s  = \int {d\omega } \frac{{\hbar k_\omega ^3 }}{{16\pi ^3 }}\left\{ {\mathop {\lim }\limits_{r \to \infty } \left\{ { - 2r^2 \left( {\int {do_r } {\bf{u}}_r } \right) \cdot \int {do_{\bf{n}} } W_\omega  \left( {\bf{n}} \right){\bf{nn}} - \int {do_r } {\bf{u}}_r \left[ {{\rm tr}{\cal Q}_\omega  \left( {{\bf{u}}_r \left| {{\bf{u}}_r } \right.} \right)} \right]} \right\}} \right\}.
\end{equation}
Because the integral of the radial unit vector over the full solid angle identically vanishes ($\int do_r \mathbf{u}_r = 0$), by using the relation ${\rm tr} \left[ {{\cal S}_\omega  \left( {{\bf{u}}_r \left| {{\bf{u}}_r } \right.} \right) - {\cal S}_\omega ^{T*} \left( {{\bf{u}}_r \left| {{\bf{u}}_r } \right.} \right)} \right] = 2i{\mathop{\rm Im}\nolimits} \left[{\rm tr}  {{\cal S}_\omega  \left( {{\bf{u}}_r \left| {{\bf{u}}_r } \right.} \right)} \right]$, we obtain
\begin{equation}
{\bf{F}}_s  = \int {d\omega } \frac{{\hbar k_\omega ^3 }}{{16\pi ^3 }}\int {do_r } {\bf{u}}_r \left\{ {\frac{{4\pi }}{{k_\omega  }}W_\omega  \left( {{\bf{u}}_r } \right){\mathop{\rm Im}\nolimits} \left[ { {\rm tr} {\cal S}_\omega  \left( {{\bf{u}}_r \left| {{\bf{u}}_r } \right.} \right)} \right] - \int {do_{\bf{n}} } W_\omega  \left( {\bf{n}} \right) {\rm tr} \left[ {{\cal S}_\omega  \left( {{\bf{u}}_r \left| {\bf{n}} \right.} \right) \cdot {\cal S}_\omega ^{T*} \left( {{\bf{u}}_r \left| {\bf{n}} \right.} \right)} \right]} \right\}
\end{equation}
Finally, by suitably relabeling the angular integration variables, it is convenient to rewrite the previous expression in the form
\begin{equation} \label{Fs}
{\bf{F}}_s \left[ {W_\omega  \left( {\bf{n}} \right)} \right] = \int {d\omega } \frac{{\hbar k_\omega ^3 }}{{16\pi ^3 }}\int {do_{\bf{n}} W_\omega  \left( {\bf{n}} \right)} \left\{ {{\bf{n}}\frac{{4\pi }}{{k_\omega  }}{\mathop{\rm Im}\nolimits} \left[ {{\rm tr}{\cal S}_\omega  \left( {{\bf{n}}\left| {\bf{n}} \right.} \right)} \right] - \int {do_{\bf{m}} } {\bf{m}}\;{\rm tr}\left[ {{\cal S}_\omega  \left( {{\bf{m}}\left| {\bf{n}} \right.} \right) \cdot {\cal S}_\omega ^{T*} \left( {{\bf{m}}\left| {\bf{n}} \right.} \right)} \right]} \right\},
\end{equation}
where we have explicitly emphasized the functional dependence of the scattering force ${\bf{F}}_s$ on the quantum weight of the squeezed vacuum $W_\omega \left( {\bf{n}} \right)$. 

\subsection{Contribution from the medium-assisted sector}
To evaluate the medium-assisted contribution ${\bf{F}}_{em}$, rather than directly computing the asymptotic behavior of the kernels ${\cal G}_{\omega \nu }$, it is highly advantageous to exploit the fundamental integral relation of Eq.(\ref{MLNF_FundIntRel}). By isolating the medium-assisted term, we can write
\begin{equation}
\sum\limits_\nu  {\int {d^3 {\bf{s}}} \,{\cal G}_{\omega \nu } \left( {\left. {\bf{r}} \right|{\bf{s}}} \right) \cdot {\cal G}_{\omega \nu }^{T*} \left( {\left. {{\bf{r}}'} \right|{\bf{s}}} \right)}  = \frac{{\hbar k_\omega ^2 }}{{\pi \varepsilon _0 }}{\mathop{\rm Im}\nolimits} \left[ {{\cal G}_\omega  \left( {\left. {\bf{r}} \right|{\bf{r}}'} \right)} \right] - \int {do_{\bf{n}} } {\cal F}_{\omega s} \left( {\left. {\bf{r}} \right|{\bf{n}}} \right) \cdot {\cal F}_{\omega s}^{T*} \left( {\left. {{\bf{r}}'} \right|{\bf{n}}} \right).
\end{equation}
Substituting this identity into the definition of ${\cal M}_{\omega em} \left( {{\bf{r}}\left| {{\bf{r}}'} \right.} \right)$ in Eq.(\ref{Dyadics_M}), the medium-assisted spectral density dyadic naturally splits into two distinct components
\begin{equation} \label{Moem1}
{\cal M}_{\omega em} \left( {{\bf{r}}\left| {{\bf{r}}'} \right.} \right) = {\cal M}_{\omega}^{(0)} \left( {{\bf{r}}\left| {{\bf{r}}'} \right.} \right) - {\cal M}_{\omega s}^{\left( 1 \right)} \left( {{\bf{r}}\left| {{\bf{r}}'} \right.} \right),
\end{equation}
where ${\cal M}_{\omega}^{(0)} \left( {{\bf{r}}\left| {{\bf{r}}'} \right.} \right)$ is the spectral density dyadic associated with the baseline isotropic fluctuations, defined as
\begin{equation}
{\cal M}_{\omega}^{(0)} \left( {{\bf{r}}\left| {{\bf{r}}'} \right.} \right) = W_{\omega em} \frac{{\hbar k_\omega ^2 }}{{\pi \varepsilon _0 }}{\mathop{\rm Im}\nolimits} \left[ {{\cal G}_\omega  \left( {\left. {\bf{r}} \right|{\bf{r}}'} \right) - \frac{1}{{k_\omega ^2 }}\nabla _{\bf{r}}  \times {\cal G}_\omega  \left( {\left. {\bf{r}} \right|{\bf{r}}'} \right) \times \mathord{\buildrel{\lower3pt\hbox{$\scriptscriptstyle\leftarrow$}} 
\over \nabla } _{{\bf{r}}'} } \right]
\end{equation}
whereas 
\begin{equation} \label{Moms1}
{\cal M}_{\omega s}^{\left( 1 \right)} \left( {{\bf{r}}\left| {{\bf{r}}'} \right.} \right) = W_{\omega em} {\mathop{\rm Re}\nolimits} \int {do_{\bf{n}} } \left[ {{\cal F}_{\omega s} \left( {\left. {\bf{r}} \right|{\bf{n}}} \right) \cdot {\cal F}_{\omega s}^{T*} \left( {\left. {{\bf{r}}'} \right|{\bf{n}}} \right) - \frac{1}{{k_\omega ^2 }}\nabla _{\bf{r}}  \times {\cal F}_{\omega s} \left( {\left. {\bf{r}} \right|{\bf{n}}} \right) \cdot {\cal F}_{\omega s}^{T*} \left( {\left. {{\bf{r}}'} \right|{\bf{n}}} \right) \times \mathord{\buildrel{\lower3pt\hbox{$\scriptscriptstyle\leftarrow$}} 
\over \nabla } _{{\bf{r}}'} } \right]
\end{equation}
is exactly the scattering dyadic ${\cal M}_{\omega s} \left( {{\bf{r}}\left| {{\bf{r}}'} \right.} \right)$ defined in Eq.(\ref{Dyadics_M}) but evaluated for an isotropic angular weight $W_{\omega s} \left( {\bf{n}} \right) = W_{\omega em}$.
By substituting the split dyadic of Eq.(\ref{Moem1}) into the expression for ${\bf{F}}_{em}$ in Eq.(\ref{Forces_s_em}), the force separates accordingly as ${\bf{F}}_{em}  = {\bf{F}}_{em}^{\left( 0 \right)}  - {\bf{F}}_{em}^{\left( 1 \right)}$ where
\begin{eqnarray}
 {\bf{F}}_{em}^{\left( 0 \right)}  &=& \varepsilon _0 \int {d\omega } \left\{ {\mathop {\lim }\limits_{r \to \infty } r^2 \int {do_r } {\bf{u}}_r  \cdot \left\{ {{\cal M}_\omega ^{\left( 0 \right)} \left( {{\bf{r}}\left| {{\bf{r}}'} \right.} \right) - \frac{1}{2}\left[ {{\rm tr}{\cal M}_\omega ^{\left( 0 \right)} \left( {{\bf{r}}\left| {{\bf{r}}'} \right.} \right)} \right]{\cal I}} \right\}_{{\bf{r}}' \to {\bf{r}}} } \right\}, \nonumber \\ 
 {\bf{F}}_{em}^{\left( 1 \right)}  &=& \varepsilon _0 \int {d\omega } \left\{ {\mathop {\lim }\limits_{r \to \infty } r^2 \int {do_r } {\bf{u}}_r  \cdot \left\{ {{\cal M}_{\omega s}^{\left( 1 \right)} \left( {{\bf{r}}\left| {{\bf{r}}'} \right.} \right) - \frac{1}{2}\left[ {{\rm tr}{\cal M}_{\omega s}^{\left( 1 \right)} \left( {{\bf{r}}\left| {{\bf{r}}'} \right.} \right)} \right]{\cal I}} \right\}_{{\bf{r}}' \to {\bf{r}}} } \right\}.
\end{eqnarray}

The term ${\bf{F}}_{em}^{\left( 0 \right)}$ represents the hypothetical thermal recoil force the object would experience if it were maintained in global thermal equilibrium within a perfectly isotropic radiation bath at its own temperature. In such a scenario, as dictated by global momentum conservation and detailed balance, an isolated body cannot undergo spontaneous self-acceleration. The momentum carried away by the emitted radiation is perfectly counterbalanced in all directions, which physically prevents self-propulsion. To rigorously prove that ${\bf{F}}_{em}^{\left( 0 \right)} = 0$, it is instructive to separate the exact dyadic Green's function in the vacuum region into its free-space and scattering contributions, ${\cal G}_\omega = {\cal G}_\omega^{{\rm fs}} + {\cal G}_\omega^{{\rm sc}}$, which correspondingly leads to the splitting of the spectral density dyadic as ${\cal M}_\omega^{(0)} = {\cal M}_\omega^{(0){\rm fs}}+{\cal M}_\omega^{(0){\rm sc}}$, thereby translating into the separation of the baseline force into two parts: ${\bf{F}}_{em}^{\left( 0 \right)} = {\bf{F}}^{{\rm fs}} + {\bf{F}}^{{\rm sc}}$. For the free-space contribution, evaluating the coincident limit ${\bf{r}}' \to {\bf{r}}$ yields the well-known result ${\mathop{\rm Im}\nolimits} [ {{\cal G}_\omega^{{\rm fs}} \left( {\left. {\bf{r}} \right|{\bf{r}}} \right)} ] = \left( {k_\omega  /6\pi } \right){\cal I}$. Including the magnetic double-curl term, the free-space spectral dyadic reduces to the purely isotropic pressure dyadic, i.e.  ${\cal M}_\omega^{(0){\rm fs}} = W_{\omega em} (\hbar k_\omega^3 / 3\pi^2 \varepsilon_0) {\cal I}$. Consequently, its angular integration over the closed spherical surface identically vanishes, yielding ${\bf{F}}^{{\rm fs}} = 0$. For the scattering contribution ${\bf{F}}^{{\rm sc}}$, we evaluate the far-field asymptotic expansion of the scattered Green's function as $r, r' \to \infty$, which takes the exact form
\begin{equation}
{\cal G}_\omega^{{\rm sc}} \left( {\left. {\bf{r}} \right|{\bf{r}}'} \right) \mathop  \approx \limits_{r,r' \to \infty } \frac{{e^{ik_\omega  \left( {r + r'} \right)} }}{{4\pi rr'}}{\cal S}_\omega  \left( {\left. {{\bf{u}}_r } \right| - {\bf{u}}_{r'} } \right).
\end{equation}
By taking the coincident limit ${\bf{r}}' \to {\bf{r}}$ (implying $r'=r$ and ${\bf{u}}_{r'} = {\bf{u}}_r$) and accounting for the magnetic contribution through the double-curl operator, the scattering spectral density dyadic is explicitly found to be
\begin{equation}
{\cal M}_\omega^{(0){\rm sc}} ({\bf{r}}|{\bf{r}}) \mathop \approx \limits_{r \to \infty} W_{\omega em} \frac{\hbar k_\omega^2}{4\pi^2 \varepsilon_0 r^2} \text{Im} \left\{ e^{2i k_\omega r} \left[ {\cal S}_\omega ({\bf{u}}_r | -{\bf{u}}_r) - {\bf{u}}_r \times {\cal S}_\omega ({\bf{u}}_r | -{\bf{u}}_r) \times {\bf{u}}_r \right] \right\}.
\end{equation}
From this expression, it is clear that the radial dependence strictly factors into highly oscillatory terms $e^{ \pm 2ik_\omega  r}/r^2$. When inserted into the force surface integral, the geometrical $1/r^2$ attenuation is exactly canceled by the spherical area measure $r^2 do_{\bf{r}}$, leaving an integral over the spectrum $d\omega$ of angular amplitudes modulated by the rapid phase $e^{ \pm 2i\omega r/c}$. By virtue of the Riemann-Lebesgue lemma, this frequency integration strictly vanishes in the limit $r \to \infty$ due to complete phase mixing, yielding ${\bf{F}}^{{\rm sc}} = 0$. This concludes the proof that ${\bf{F}}_{em}^{\left( 0 \right)} = 0$.

The remaining term ${\bf{F}}_{em}^{\left( 1 \right)}$, by virtue of the expression for ${\cal M}_{\omega s}^{\left( 1 \right)}$ given in Eq.(\ref{Moms1}), can be written as ${\bf{F}}_{em}^{\left( 1 \right)} = {\bf{F}}_s[W_{\omega em}]$, where ${\bf{F}}_s[W_{\omega em}]$ is, from Eq.(\ref{Fs}), the scattering contribution to the force evaluated at $W_{\omega s} \left( {\bf{n}} \right) = W_{\omega em}$. Physically, the term ${\bf{F}}_{em}^{\left( 1 \right)}$ represents the hypothetical scattering force the object would experience if it were externally illuminated by a perfectly isotropic radiation bath of intensity $W_{\omega em}$. By virtue of Kirchhoff's law of thermal radiation, the directional emissivity of an object is fundamentally dictated by its directional absorptivity. Consequently, the actual net momentum lost by the body due to its own anisotropic thermal emission ($-{\bf{F}}_{em}^{\left( 1 \right)}$) is exactly equal and opposite to this hypothetical scattering force. Since the thermodynamic weight $W_{\omega em}$ is isotropic and thus independent of the angular variable ${\bf{n}}$, it factors out of the integration. Utilizing our previously derived expression for ${\bf{F}}_s$, we immediately obtain the exact expression for the medium-assisted recoil force
\begin{equation}
{\bf{F}}_{em}  =  - \int {d\omega } \frac{{\hbar k_\omega ^3 }}{{16\pi ^3 }}W_{\omega em} \int {do_{\bf{n}} } \left\{ {{\bf{n}}\frac{{4\pi }}{{k_\omega  }}{\mathop{\rm Im}\nolimits} \left[ {{\rm tr}{\cal S}_\omega  \left( {{\bf{n}}\left| {\bf{n}} \right.} \right)} \right] - \int {do_{\bf{m}} } {\bf{m}}\;{\rm tr}\left[ {{\cal S}_\omega  \left( {{\bf{m}}\left| {\bf{n}} \right.} \right) \cdot {\cal S}_\omega ^{T*} \left( {{\bf{m}}\left| {\bf{n}} \right.} \right)} \right]} \right\}.
\end{equation}

\section{Electromagnetic scattering from a homogeneous sphere}
In this appendix, we summarize the standard Mie theory \cite{Krist2016, Bornn2019, Bohre1983} and the corresponding optical cross-sections for the electromagnetic scattering by a non-magnetic, homogeneous, isotropic lossy sphere of radius $a$ and complex permittivity $\varepsilon_\omega^V$ (see Appendix A), which are employed in Sec.IVB. The wave number inside the material is $k_\omega^V = k_\omega \sqrt{\varepsilon_\omega^V}$. The scattering dyadic ${\cal S}_\omega({\bf{m}}|{\bf{n}})$ can be expanded in the basis of the standard vector spherical harmonics ${\bf{X}}_{nm}({\bf{m}})$ as \cite{Ciatb2025}
\begin{equation} \label{D_Somn}
{\cal S}_\omega({\bf{m}}|{\bf{n}}) = \frac{4\pi i}{k_\omega} \sum_{n=1}^\infty \sum_{m=-n}^n \left[ a_{n\omega} {\bf{m}} \times {\bf{X}}_{nm}({\bf{m}}) {\bf{n}} \times {\bf{X}}_{nm}^*({\bf{n}}) + b_{n\omega} {\bf{X}}_{nm}({\bf{m}}) {\bf{X}}_{nm}^*({\bf{n}}) \right].
\end{equation}
The Mie scattering coefficients $a_{n\omega}$ and $b_{n\omega}$, which characterize the excitation of the transverse magnetic (TM) and transverse electric (TE) modes, respectively, are given by
\begin{align}
a_{n\omega} &= \frac{\displaystyle j_n(k_\omega^V a) \frac{\partial}{\partial a}[a j_n(k_\omega a)] - \frac{1}{\varepsilon_\omega^V} j_n(k_\omega a) \frac{\partial}{\partial a}[a j_n(k_\omega^V a)]}{\displaystyle j_n(k_\omega^V a) \frac{\partial}{\partial a}[a h_n^{(1)}(k_\omega a)] - \frac{1}{\varepsilon_\omega^V} h_n^{(1)}(k_\omega a) \frac{\partial}{\partial a}[a j_n(k_\omega^V a)]}, &
b_{n\omega} &= \frac{\displaystyle j_n(k_\omega^V a) \frac{\partial}{\partial a}[a j_n(k_\omega a)] - j_n(k_\omega a) \frac{\partial}{\partial a}[a j_n(k_\omega^V a)]}{\displaystyle j_n(k_\omega^V a) \frac{\partial}{\partial a}[a h_n^{(1)}(k_\omega a)] - h_n^{(1)}(k_\omega a) \frac{\partial}{\partial a}[a j_n(k_\omega^V a)]},
\end{align}
where $j_n(z)$ are the spherical Bessel functions of the first kind and $h_n^{(1)}(z)$ are the spherical Hankel functions of the first kind. The macroscopic scalar optical cross-sections \cite{Bohre1983} can be formally derived from specific evaluations and angular integrals involving the trace of the scattering dyadic. Because the trace operator acts within the two-dimensional transverse plane, it performs an unpolarized summation over the two independent orthogonal polarization states. This summation inherently introduces a factor of $2$ when mapping the dyadic traces to the standard unpolarized scalar cross-sections. Specifically, the extinction cross-section is determined via the generalized optical theorem by evaluating the imaginary part of the dyadic trace in the exact forward scattering direction (${\bf{m}} = {\bf{n}}$). Analogously, the total scattering cross-section is obtained by integrating the differential scattered intensity, represented by the trace of the dyadic norm, over all outgoing solid angles:
\begin{align}
2 \sigma_\omega^{ext} &= \frac{4\pi}{k_\omega} \mathop{\rm Im}\nolimits \left[ {\rm tr} {\cal S}_\omega({\bf{n}}|{\bf{n}}) \right], &
2 \sigma_\omega^{sca} &= \int do_{\bf{m}} {\rm tr} \left[ {\cal S}_\omega({\bf{m}}|{\bf{n}}) \cdot {\cal S}_\omega^{T*}({\bf{m}}|{\bf{n}}) \right].
\end{align}
By substituting the multipole expansion of the scattering dyadic in Eq.(\ref{D_Somn}), these trace relations yield the standard Mie series:
\begin{align}
\sigma_\omega^{ext} &= \frac{2\pi}{k_\omega^2} \sum_{n=1}^\infty (2n+1) \left[ \mathop{\rm Re}\nolimits(a_{n\omega}) + \mathop{\rm Re}\nolimits(b_{n\omega}) \right], &
\sigma_\omega^{sca} &= \frac{2\pi}{k_\omega^2} \sum_{n=1}^\infty (2n+1) \left[ |a_{n\omega}|^2 + |b_{n\omega}|^2 \right].
\end{align}
The absorption cross-section follows as $\sigma_\omega^{abs} = \sigma_\omega^{ext} - \sigma_\omega^{sca}$. Furthermore, the asymmetry cross-section $\sigma_\omega^{asym} = \sigma_\omega^{sca} \langle \cos\theta \rangle$, which weights the scattered intensity by the angular projection factor ${\bf{n}} \cdot {\bf{m}} = \cos\theta$, satisfies the trace relation
\begin{equation}
2 \sigma_\omega^{asym} = \int do_{\bf{m}} ({\bf{n}} \cdot {\bf{m}}) \, {\rm tr} \left[ {\cal S}_\omega({\bf{m}}|{\bf{n}}) \cdot {\cal S}_\omega^{T*}({\bf{m}}|{\bf{n}}) \right],
\end{equation}
which evaluates to the analytical series
\begin{equation}
\sigma_\omega^{asym} = \frac{4\pi}{k_\omega^2} \sum_{n=1}^\infty \left\{ \frac{n(n+2)}{n+1} \mathop{\rm Re}\nolimits \left[ a_{n\omega} a_{(n+1)\omega}^* + b_{n\omega} b_{(n+1)\omega}^* \right] + \frac{2n+1}{n(n+1)} \mathop{\rm Re}\nolimits \left( a_{n\omega} b_{n\omega}^* \right) \right\}.
\end{equation}
Finally, the radiation pressure cross-section $\sigma_\omega^{pr}$, which dictates the net momentum transfer from the incident field to the spherical object, is given by
\begin{equation}
\sigma_\omega^{pr} = \sigma_\omega^{ext} - \sigma_\omega^{asym} \geq 0.
\end{equation}

\end{document}